\documentclass[12pt, draftclsnofoot, onecolumn]{IEEEtran}

\usepackage{float,color,comment}
\usepackage{amsmath}
\usepackage{caption}
\usepackage{subfigure}
\usepackage{graphics} 
\usepackage{epsfig} 

\usepackage{amsfonts,amssymb}
\usepackage{multirow}
\graphicspath{{figure/}}

\usepackage{mathtools, nccmath}

\DeclarePairedDelimiter{\nint}\lfloor\rceil

\DeclareMathOperator*{\argminA}{arg\,min} 

\begin{document}
%
\title{\LARGE \bf
An Efficient Deep Learning Framework for Low Rate Massive MIMO CSI Reporting}
%
%
%

\author{Zhenyu~Liu,
Lin~Zhang, and Zhi~Ding
\thanks{Z. Liu is with the School of Information and Communication Engineering,
Beijing University of Posts and Telecommunications, Beijing 100876, China,
and also with the Department of Electrical and Computer Engineering,
University of California at Davis, Davis, CA 95616 USA (e-mail:
lzyu@bupt.edu.cn).}
\thanks{L. Zhang is with the School of Information and Communication
Engineering, Beijing University of Posts and Telecommunications, Beijing
100876, China (e-mail: zhanglin@bupt.edu.cn).}
\thanks{Z. Ding is with the Department of Electrical and Computer Engineering,
University of California at Davis, Davis, CA 95616 USA (e-mail:
zding@ucdavis.edu).}}

\maketitle

\begin{abstract}

Channel state information (CSI) reporting is important for 
multiple-input multiple-output (MIMO) transmitters to achieve 
high capacity and energy efficiency in frequency division duplex (FDD) mode. 
CSI reporting for massive MIMO systems could consume excessive
bandwidth and degrade spectrum efficiency. 
Deep learning (DL)-based compression integrated with channel correlations 
have demonstrated success in improving CSI recovery. 
However, existing works focusing on CSI compression have shown little
on the efficient encoding of CSI report. In this paper, we propose an efficient DL-based compression framework (called CQNet) to jointly tackle CSI compression, report
encoding, and recovery under bandwidth constraint. 
CQNet can be  directly integrated within other  DL-based  CSI feedback works for further
enhancement. CQNet significantly outperforms solutions using uniform CSI 
quantization and $\mu$-law non-uniform quantization. Compared with
traditional CSI reporting, much fewer bits are required to 
achieve comparable CSI reconstruction accuracy.


\end{abstract}


%
\IEEEpeerreviewmaketitle

\section{Introduction}
Massive multiple-input multiple-output (MIMO) systems have shown great promise in
delivering high
spectrum and energy efficiency for 5G
and future wireless communication systems \cite{Massive_MIMO}. 
By utilizing a large number
of antennas in massive MIMO framework, gNB (or gNodeB) in 5G 
can achieve very high downlink throughput if sufficiently accurate downlink
channel state information (CSI) is available at the gNB.
Consequently, gNB needs to acquire the downlink CSI in an accurate and timely manner
to fully utilize 
the spatial diversity and multiplexing gains. 

In time division duplex (TDD) systems, gNB can leverage its uplink CSI as 
the close estimate of its downlink CSI based on the well
known reciprocity between downlink and uplink CSIs.
In frequency division duplex (FDD) systems, however, uplink and downlink channels are in different
frequency bands. Thus, it is difficult to only
rely on uplink CSI to estimate the downlink CSI as 
the bi-directional channel reciprocity no longer applies.
Consequently,
gNB transmitters of FDD systems would require user equipment (UE)
to provide certain CSI reporting about the downlink CSI. For massive MIMO,
such feedback data can be substantial since the large number
of antennas leads to very high CSI dimensionality. 
Large bandwidth in high rate links further exacerbates the high
feedback load.

To improve spectrum efficiency for CSI reporting in FDD systems, 
compressed sensing (CS)-based approaches can exploit the CSI
properties of low rank or sparsity
to derive a compressed CSI representation for feedback.  
Two major correlation properties used in CS-based CSI feedback approaches
include spatial CSI correlation \cite{cs2,cs4_3,cs5_3} that stems from the limited scattering characteristics of signal propagation, and temporal CSI correlation \cite{cs3_3} 
owing to Doppler effects. However, CS-based approaches still have some limitations.
On the one hand, CS-based approaches require a strong channel sparsity condition which is not strictly held in some cases. On the other hand, CS algorithms 
are often iterative and computationally intensive during decoding procession,
which may lead to long delays. 

Deep learning (DL) is a powerful tool for exploring the underlying structures 
from large data set, and has been widely used in computer vision and natural language processing. It can play a helpful role in CSI estimation when traditional methods 
generate limited performance. 
There have been some successful applications to derive
reliable downlink CSI in massive MIMO systems for
channel estimation \cite{dl1} and low rate CSI feedback \cite{dls,dlo}.  
In particular for massive MIMO, 
the authors of \cite{dls} developed a CSI compression and 
recovery mechanism using an autoencoder structure \cite{autoencoder3}, 
and demonstrated better accuracy than CS-based methods 
in terms of downlink CSI reconstruction from limited UE feedback. 
The work of \cite{dlo} further exploited the FDD bi-directional channel correlation.
By jointly utilizing the available uplink CSI and low rate UE feedback in massive
MIMO systems at gNB to recover the
unknown downlink CSI, significant performance gain of downlink CSI estimate was shown 
over DL architecture based only on UE feedback \cite{dlo}.

Common implementation of DL networks complies with IEEE 754 standard \cite{float_standard}. Single precision, which is defined in this standard and used in \cite{dls, dlo}, is the most typical data type adopted in DL networks \cite{float1}. 
Although DL methods have delivered noticeable performance improvement 
in reducing the dimension of CSI matrices in massive MIMO, 
current CSI reporting results that
use single precision (32-bit) to encode feedback coefficients 
still consume too much bandwidth. 
To reduce the number of bits required by each codeword, low-bit quantization
should take place in encoding after dimension compression. 
Clearly, low-bit quantization for CSI codewords will degrade CSI reconstruction accuracy. 
Therefore, it is important to jointly optimize the encoding
and dimension compression to maintain high CSI reconstruction
accuracy without consuming excessive bandwidth in FDD systems. 

In this work, we propose an efficient CSI compression solution
and design an end-to-end DL framework CQNet to optimize downlink
CSI compression and feedback encoding simultaneously. 
Our CQNet is a simple plug-in that can be integrated into
existing DL-based CSI feedback frameworks. We also 
investigate how feedback bandwidth affects CSI reconstruction accuracy 
under a uniform quantization framework. Moreover, we use
two existing works CsiNet \cite{dls} and DualNet \cite{dlo} 
as examples to show how to implement the CQNet to improve bandwidth efficiency.
To enhance the flexibility of phase quantization in DualNet, 
we also propose a DL-based magnitude-adaptive phase encoder, 
which can easily adjust quantization bits according to required accuracy. 

Our contributions in this paper are summarized as follows:

\begin{itemize}
    \item We evaluate how CSI reconstruction accuracy is affected by the
    feedback bandwidth under uniform quantization, and show
    significant bandwidth  savings over single precision feedback of compressed CSI coefficients.
     
    \item We develop a CQNet that can simultaneously optimize 
    the dimension compression and encoding of downlink CSI to improve the
    bandwidth efficiency.  
The CQNet can be directly combined with existing DL-based CSI feedback 
methods to save bandwidth.

\item We analyze different encoding and quantization codewords to demonstrate the
advantages of CQNet.

\item We further design a specialized DL-based phase quantization framework that
can achieve the magnitude-adaptive phase quantization. 
This framework increases the flexibility of phase quantization
and can regulate the weight of quantization entropy
to achieve a balanced bandwidth versus CSI accuracy design trade-off. 

\item Test results demonstrate that CQNet significantly outperforms uniform quantization and $\mu$-law quantization. Compared with single-precision feedback, CQNet can achieve
comparable CSI reconstruction accuracy using 5 bits per coefficient
in the feedback. CQNet with  
entropy encoding can further reduce down to 4 bits per coefficient.

\end{itemize}

\section{Related Works}

FDD base stations like eNB or gNB typically rely on CSI feedbacks to
report downlink CSI due to the weak reciprocity
between uplink and downlink channels. 
To implement massive MIMO downlink in FDD, 
feedback payload is substantially high because 
of the large antenna number and
wide bandwidth. Conventional
methods based purely on UE feedback face several challenges
including channel models and feedback bandwidth consumption.
Recognizing the importance
to conserve feedback bandwidth and improve downlink CSI
reconstruction accuracy,
there has been a recent surge of 
interest in DL-based CSI feedback \cite{dls,dlo, dllstm, dlrnn}.
\begin{figure}[thpb]
      \centering
    
      \includegraphics[scale=0.4]{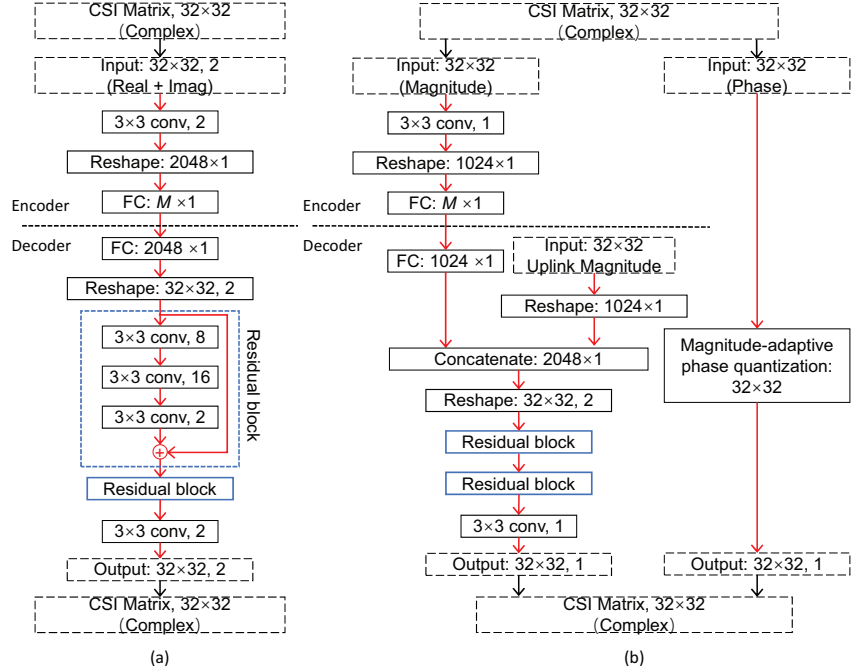}
      \caption{Architecture of CsiNet and DualNet-MAG. CsiNet consists of an encoder network with a 2-channel 3 $\times$ 3 convolutional (conv) layer and an $M \times 1$ fully connected layer (FC) for dimension compression and a decoder using a FC for dimension decompression and two Residual blocks for CSI calibration. DualNet-MAG concatenates the feedback codewords for magnitude with uplink CSI's magnitude before feeding into the residual blocks. The phase of CSI is fed back using  the magnitude-adaptive quantization.}
      \label{figurecd}
  \end{figure}

Among related works, a DL-based CSI feedback framework CsiNet for massive MIMO downlink
was proposed in \cite{dls} to 
reduce UE feedback overhead. As shown in Fig. \ref{figurecd} (a), CsiNet utilizes 
an autoencoder architecture, where an encoder deep neural network (DNN) acts as
a compression module and a corresponding
decoder DNN is responsible for CSI reconstruction. 
Treating CSI matrix as a virtual image, 
convolutional layer is used in both encoder
and decoder to exploit its spatial and spectral correlation. 
Fully connected layers are used for dimension compression and decompression. 
Each CSI matrix is split into real and imaginary parts, rearranged into two 
sets of encoder DNN input. CsiNet demonstrated performance gain over
CS-based methods.

Another DL-based CSI feedback framework DualNet-MAG established
the bi-directional correlation among the magnitudes of FDD channel 
coefficients \cite{dlo}. DualNet-MAG demonstrated significant 
performance benefit of exploiting this uplink-downlink CSI correlation in reducing the 
amount of CSI feedback \cite{dlo}.
Fig. \ref{figurecd} (b) shows that, unlike CsiNet, 
DualNet exploits the available uplink CSI at gNB to 
help recover the downlink CSI from lower rate CSI feedback.  
The bi-directional channel correlation between CSI magnitudes 
helps the decoder DNN recover the downlink CSI magnitudes with
better accuracy by leveraging the low-rate feedback codewords 
and locally available uplink CSI magnitudes. The work in  \cite{dlo}
also shows that the CSI phase correlation is
weak between uplink and downlink CSIs. Thus, CSI feedback 
should focus on CSI phase encoding and a magnitude-adaptive phase
encoding can reduce the phase feedback overhead
and achieve clear performance improvement over traditional methods and 
CsiNet. 

Another work \cite{dllstm} proposed the use of long-short time
memory (LSTM) network \cite{LSTM} known as CsiNet-LSTM to exploit the temporal correlation of CSI.
In \cite{dlrnn}, a new LSTM network further reduced the 
number of parameters to be trained while maintaining the CSI recovery accuracy.

However, the aforementioned works on DL-based CSI feedback generally focused on 
dimension reduction of CSI feedback. 
Single precision or float32 (defined in IEEE 754 standard) is the most typical data type adopted in DL networks \cite{float1} and has been widely used in channel estimation \cite{dl1} and channel feedback \cite{dls,dlo}. Although single precision can help DL networks gain great accuracy,
bandwidth efficient encoding of dimension-reduced 
CSI coefficients while maintaining the comparable accuracy remains a challenge.

A recent work \cite{float2} showed that encoding precision can be reduced with 
manageable accuracy loss in image classification. 
To achieve better CSI reporting efficiency, we investigate
the encoding of compressed CSI coefficients
for limited feedback bandwidth with minimum loss of CSI reconstruction accuracy.
In this paper, we first evaluate the impact of feedback bandwidth on CSI 
reconstruction accuracy under uniform quantization/encoding, 
and show that quantization can significantly save the bandwidth. 
Based on this observation, we further propose a novel learning framework to
jointly optimize the dimension compression and CSI encoding
at the same time. This framework can reduce the bandwidth 
required by downlink CSI feedback while preserving
high CSI reconstruction accuracy.
It can also be easily integrated with the existing DL-based CSI feedback works.

There are a few other recent works involved the quantization of the CSI codewords. In \cite{guo2019convolutional},  the  $\mu$-law quantization was utilized to encode the codewords after dimension compression. However, neither $\mu$-law quantization 
nor uniform quantization alone
can efficiently encode CSI feedbacks, which we shall
demonstrate in the performance evaluation section of
this work. 
Another recent method \cite{Quan_1} proposed
a specific quantization DL network with successful results.
Advancing further, our framework in this work
can customize quantization parameters for 
each CSI codeword and can be easily integrated into
 existing CSI feedback works without additional 
modifications. 

\section{System Model}
Consider a single-cell massive MIMO system, in which
the gNB has $N_b \gg 1$ antennas and UEs have a single antenna. 
The system applies orthogonal frequency division multiplexing (OFDM) 
over $N_f$ subcarriers, for which the downlink received signal at the $n-$th subcarrier is\vspace*{-2mm}
\begin{equation}
	y_{d}^{(n)} ={\mathbf{h}_d^{(n)}}^H\mathbf{w}_T^{(n)}x_d^{(n)} + n_d^{(n)}, 
\vspace*{-1mm}\label{equ1}
\end{equation}
where $\mathbf{h}_d^{(n)} \in \mathbb{C}^{N_b\times1}$ denotes the channel 
vector of the $n-$th subcarrier, $\mathbf{w}_T^{(n)} \in \mathbb{C}^{N_b\times1}$ denotes transmit beamformer, $x_d^{(n)}\in \mathbb{C}$ is the transmitted symbol, 
and $n_d^{(n)}\in \mathbb{C}$ denotes the additive noise. 
$(\cdot)^H$ denotes conjugate transpose. With the downlink channel vector $\mathbf{h}_d^{(n)}$, gNB can calculate the transmit beamformer $\mathbf{w}_T^{(n)}$.
The uplink received signal of the $n-$th subcarrier is given by\vspace*{-2mm}
\begin{equation}
	y_{u}^{(n)} ={\mathbf{w}_R^{(n)}}^H\mathbf{h}_u^{(n)}x_u^{(n)} + {\mathbf{w}_R^{(n)}}^H\mathbf{n}_u^{(n)}, \label{equUL}
\end{equation}
where $\mathbf{w}_R^{(n)} \in \mathbb{C}^{N_b\times1}$ denotes the receive beamformer, and subscript $u$ 
denotes uplink signals and noise, similar to (\ref{equ1}).
The downlink and uplink CSI matrices in the spatial frequency domain 
are denoted as $\tilde{\mathbf{H}}_d = \left[\mathbf{h}_d^{(1)},..., \mathbf{h}_d^{(N_f)}\right]^H \in \mathbb{C}^{N_f\times N_b}$ and $\tilde{\mathbf{H}}_u = \left[\mathbf{h}_u^{(1)},..., \mathbf{h}_u^{(N_f)}\right]^H \in \mathbb{C}^{N_f\times N_b}$, respectively.  
%


To reduce the feedback overhead, we first exploit the property
that CSI matrices exhibit
some sparsity in the delay domain since the delay among
multiple paths lies in a particularly limited period
\cite{sparse_3}. The CSI matrix 
$\mathbf{H}_f$ in 
frequency domain can be transformed  to be $\mathbf{H}_t$ in delay domain  using an inverse discrete Fourier transform (IDFT), i.e.,
\begin{equation}
	\mathbf{H}_f \mathbf{F}^H = \mathbf{H}_t, \label{idft3}
\end{equation}%
where $\mathbf{F}$ and $\mathbf{F}^H $ denote the $N_f \times N_f$ unitary DFT matrix
and IDFT matrix, respectively.  
After IDFT of (\ref{idft3}), most elements in the $N_f\times N_b$
matrix $\mathbf{H}_t$ are near zero except for the
first $Q_f$ rows. Therefore, we truncate the
channel matrix to the first $Q_f$ rows that are with 
distinct non-zero values, and utilize $\mathbf{H}_d$ and $\mathbf{H}_u$ to denote the first $Q_f$ rows of matrices after IDFT of $\tilde{\mathbf{H}}_d$ and $\tilde{\mathbf{H}}_u$, respectively. 

To reduce the redundancy in reporting downlink CSI,  
the codewords should be dimension-compressed and low-bit encoded. 
Consequently, unlike CsiNet and DualNet-MAG which only utilize 
the encoder and decoder for CSI dimension compression and reconstruction respectively, 
a quantizer module is added between the encoder and decoder to execute
dimension compression and encoding/quantization jointly. 

\begin{figure}
    \centering
    \includegraphics[scale=0.5]{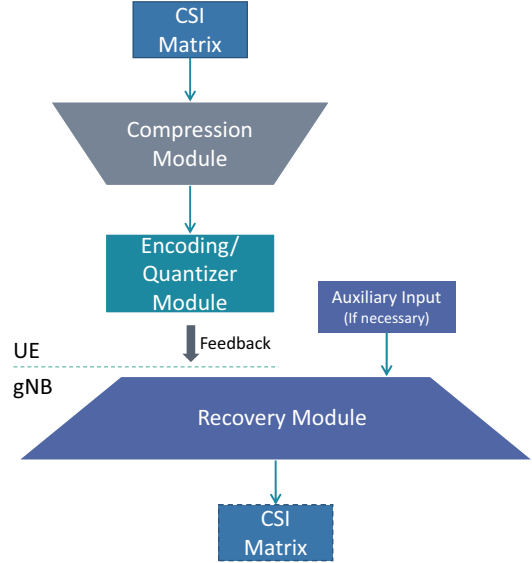}\vspace*{-1mm}
    \caption{CSI feedback framework.}
    \label{figNFIG1}
\end{figure}
  
As shown in Fig.~\ref{figNFIG1},  our CSI feedback framework for FDD
downlink channel reconstruction consists of 
3 modules: a compression  module,  an encoding/quantizer module, and a 
recovery module. 
Specifically, we design a DNN architecture CQNet to jointly optimize the dimension
compression and encoding process at the receiver. The recovery module at 
the transmitter is correspondingly optimized for CSI recovery. 
We shall illustrate that the CQNet framework can be integrated 
with existing dimension compression 
methods such as CsiNet in \cite{dls} and DualNet-MAG in \cite{dlo}.
Jointly optimizing compression and encoding of downlink CSI feedback,
these corresponding new architectures are respectively named as CsiQnet and DualQnet. 

We shall
let $\mathbf{\hat{H}}_d$ denote the reconstructed downlink 
CSI matrix. 
We define the encoding/quantization function as $f_{\rm quan}(\cdot)$.
For CsiQnet, the dimension compression module, quantizer module, decoder module can be denoted, respectively, by 
\begin{eqnarray}
\qquad\qquad	\mathbf{s}_1 &=& f_{c,1}(\mathbf{H}_d),  \\
\hat{\mathbf{s}}_1 &=& f_{\rm quan,1}(\mathbf{s}_1),
\\
	\hat{\mathbf{H}}_{d} &=& f_{r,1}(\hat{\mathbf{s}}_1).
\end{eqnarray}%
For DualQnet, the dimension compression module, quantizer module, decoder module can be denoted, respectively, by 

\begin{eqnarray}
\qquad\qquad	\mathbf{s}_2 &=& f_{c,2}(\mathbf{H}_d),  \\
\hat{\mathbf{s}}_2 &=& f_{\rm quan,2}(\mathbf{s}_2),
\\
	\hat{\mathbf{H}}_d &=& f_{r,2}(\hat{\mathbf{s}}_2,\mathbf{H}_u).
\end{eqnarray}%
The optimization of downlink CSI recovery method can be formulated as minimizing $\left \|  \mathbf{H}_d - \hat{\mathbf{H}}_d\right \|^2$, where $\left \|  \cdot\right \|$ is Frobenius norm.



\section{Bandlimited CSI feedback}

The DL-based CSI feedback works including CsiNet in \cite{dls} and 
DualNet in \cite{dlo} have demonstrated substantial performance gain
in terms of downlink CSI feedback reduction and reconstruction accuracy. 
However, in addition to the benefit of downlink CSI compression,
additional encoding of the compressed CSI feedback coefficients from the
original float32 format \cite{dls,dlo} can further
reduce the downlink CSI feedback for massive MIMO systems drastically. 

Toward this goal, we first evaluate the impact of quantization/encoding 
codeword length on CSI reconstruction accuracy by
examining a simple uniform quantizer. 
We start by testing CsiNet and DualNet by simply adding the uniform quantizer
between the encoder DNN and decoder DNN of CsiNet and DualNet 
to assess the impact of quantization distortion.

For CsiNet, Fig.~\ref{figurecd} (a) shows the encoder network 
containing a $3 \times 3$ convolutional layer with 2 channels 
plus an $M$-unit fully connected layer for dimension compression.
The decoder network consists of a fully connected layer for decompression 
and two residual  blocks to reconstruct the downlink CSI. 
Each residual  block contains three $3 \times 3$ convolutional layers 
with the channel number 8, 16, and 2, respectively. 

DualNet-MAG leverages the magnitude correlation between uplink and downlink  to improve 
CSI feedback efficiency. As shown in Fig.~\ref{figurecd} (b), DualNet-MAG 
processes the magnitude and phase separately. After separation, the CSI magnitudes
are sent to the encoder network including  a $3 \times 3$ convolutional layer and an $M$-unit fully connected layer. 
The gNB decoder receives the compressed codewords and 
uses the locally available uplink CSI magnitudes 
together to jointly decode downlink CSI. 
The received codewords are first mapped to their 
original length using a fully connected layer.
The conjugation layer combines both downlink CSI 
and uplink CSI to generate an output reshaped into $2$ 
feature maps to be used as input to the residual blocks. 
To save feedback bandwidth while limiting quantization error, a
magnitude-adaptive phase quantization (MAPQ) 
is applied in which CSI coefficients with larger magnitude
receive finer phase quantization, and vice versa. 

Uniform quantization is simple and well known in practice. 
It is basically a rounding process, in which each sample value is rounded 
to the nearest value among a finite set of possible quantization levels.
We can normalize to limit the amplitude of CSI coeffients 
between $[s_{\rm min},s_{\rm max}]$. Let $\ell$ be the number of bits
for amplitude quantization. Each CSI coefficient's amplitude can be
uniformly quantized into $2^\ell$ levels:
\begin{equation}
\hat{s}  = \Delta \nint{\frac{s}{\Delta}},\qquad  \mbox{where}\quad
\Delta = \frac{s_{\rm max}-s_{\rm min}}{2^\ell - 1}.
\end{equation}
We include uniform quantization into the DL-based CSI feedback framework. 
Specifically, we first train CsiNet and DualNet-MAG without quantization in
the original end-to-end approach. Next, we use uniform quantization to 
digitize the compressed CSI coefficients, before sending the quantized 
CSI values into the decoder. 

\begin{figure} 
	\hspace*{-4mm}
\subfigure[CsiNet] {\label{figeq:a} 
\includegraphics[width=0.5\columnwidth]{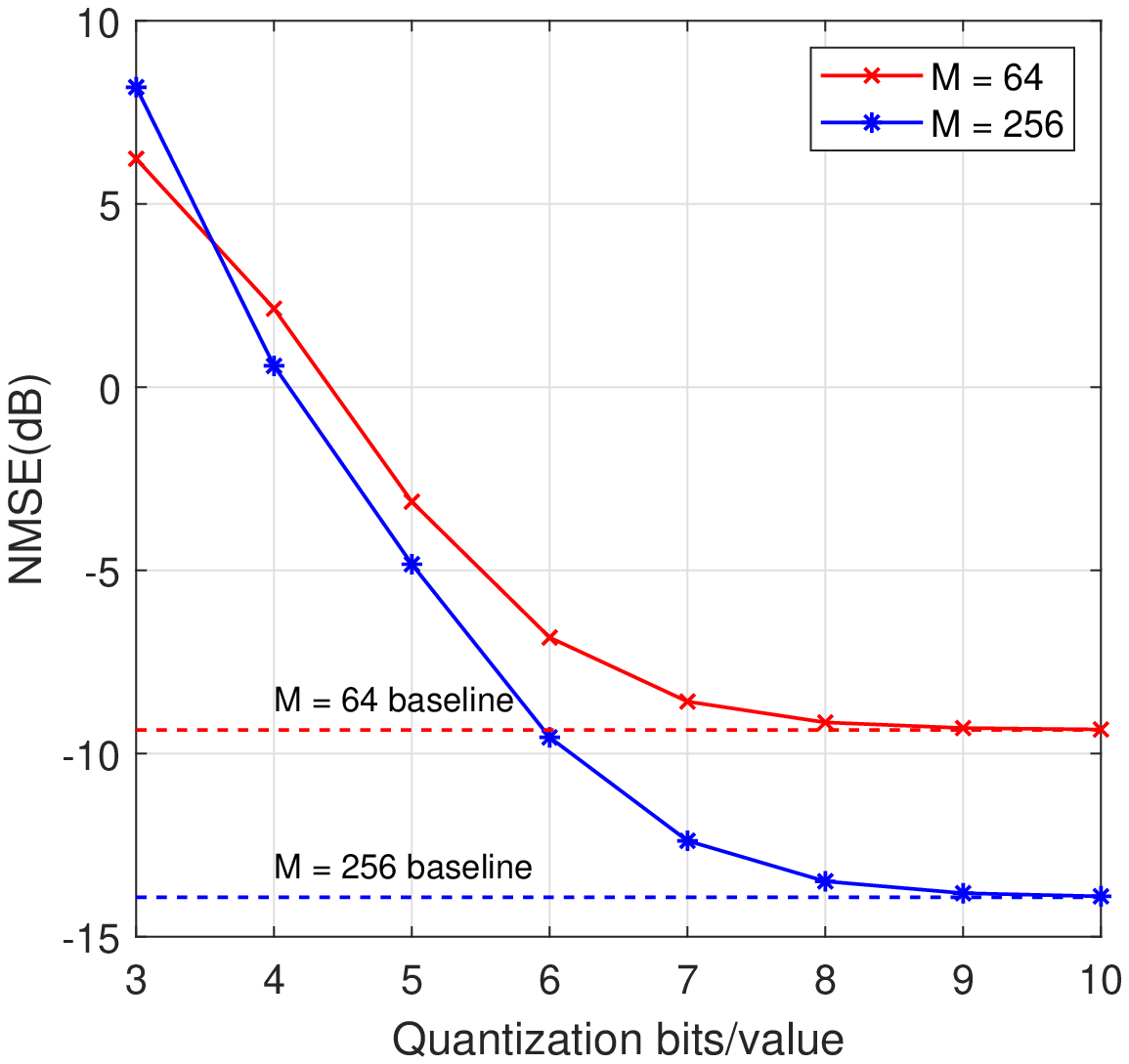}
} \hspace*{-5mm}
\subfigure[DualNet-MAG] { \label{figeq:b} 
\includegraphics[width=0.5\columnwidth]{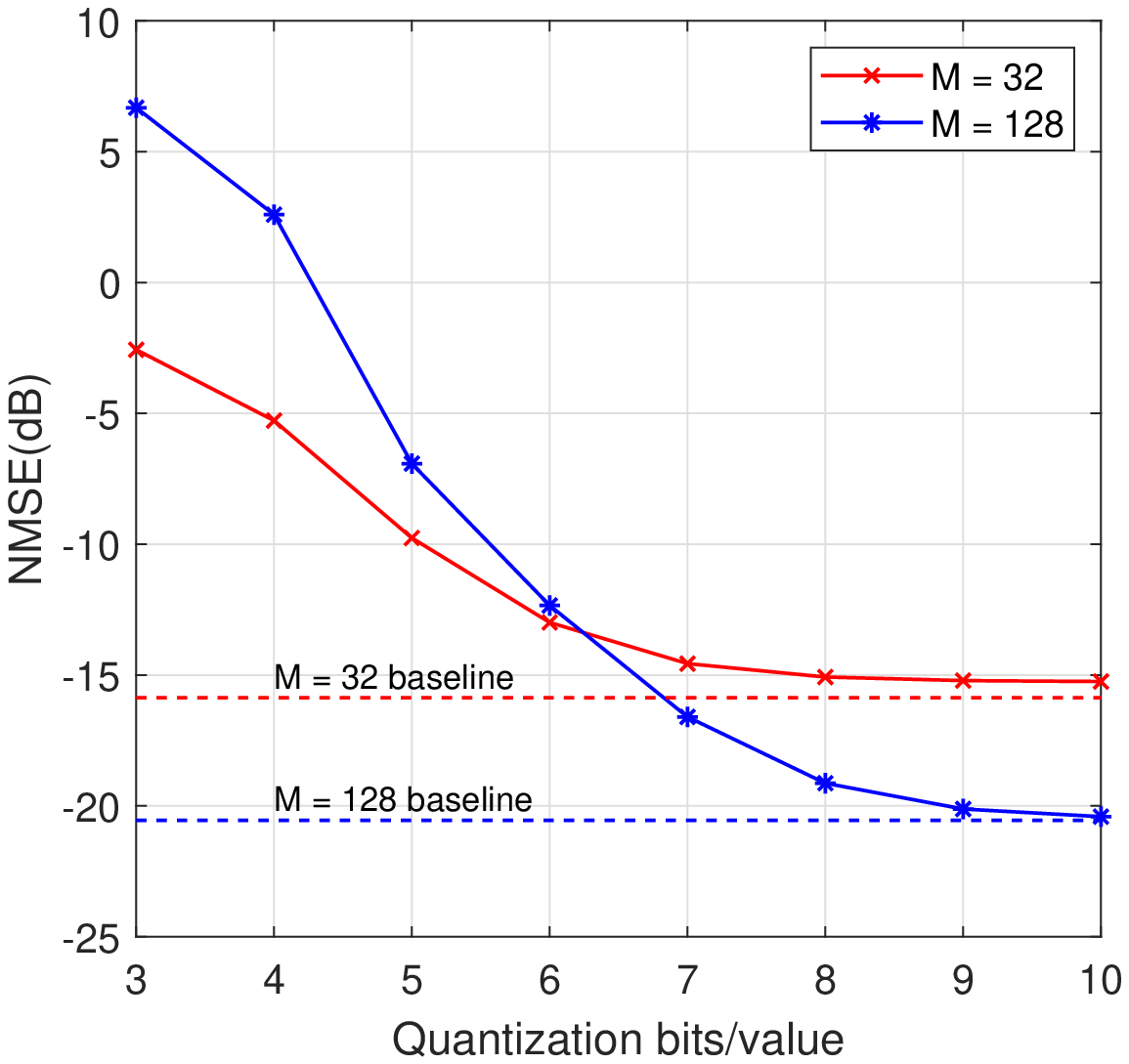} 
} \vspace*{-2mm}
\caption{CSI recovery accuracy at different quantization levels.} 
\label{figureeq} \vspace*{-3mm}
\end{figure}

In the next experiment,  we use the COST 2100 channel model to generate 
CSI matrices for reconstruction  evaluation \cite{c2100}. A uniform linear 
array (ULA) of transmit antennas is set up with half-wavelength spacing in an 
indoor environment with uplink and downlink bands 
at $5.1$ GHz and $5.3$ GHz, respectively. 
Fig. \ref{figureeq} shows the resulting NMSE for CsiNet and DualNet-MAG 
under different levels of quantization $2^\ell$ as we vary
$\ell$ from $3$ to $10$ bits. 
For CsiNet, we set the length of compressed codeword vector to $M = 64$ and $256$,
respectively. For DualNet-MAG, we set the length of compressed codeword 
vector to $M = 32$ and $128$, respectively. 
Results from float32 serve as the baseline in CSI feedback.

A few observations can be made from the results of Fig. \ref{figureeq}. 
First, high precision feedback of $32$ bits per CSI value 
is unnecessary, as 10-bit uniform quantizer achieves nearly the same 
accuracy as float32 for both CsiNet and DualNet. 
Second, both CsiNet and DualNet are more robust to quantization errors at lower
compression. Third, DualNet generally achieves better accuracy. 
Finally,  CSI reconstruction accuracy degrades with 
coarser quantization. When $\ell$ drops below $7$, the 
reconstruction experiences a clear degradation. 

These test results motivate our study to design a more efficient and
suitable encoding/quantization solution which can deliver high CSI reconstruction 
accuracy while using smaller $\ell$ to further improve the feedback 
bandwidth efficiency.


\section{CQNet}

Optimum encoding or quantization depends on the distribution
of data under quantization and the performance metric. For example, 
non-uniform quantizer such as the $\mu$-law method improved
the signal-to-quantization noise ratio (SQNR) for lower power
signals. Clearly, it is impractical to exhaustively test a huge number
of encoding schemes for the downlink CSI 
coefficients generated by the encoder DNN in order to 
determine the best fit. Instead, we shall develop a 
novel DNN approach to learn and optimize
the quantization intervals in order to improve the CSI
recovery accuracy for limited number of quantization levels.

\subsection{Joint Compression and Quantization Encoding}

\begin{figure}
    \centering
    \includegraphics[scale=0.45]{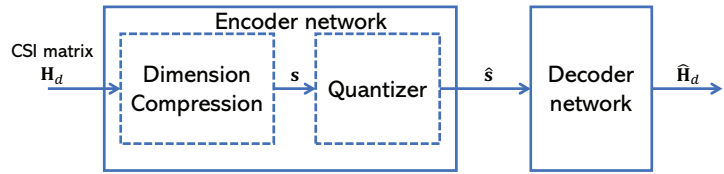}\vspace*{-1mm}
    \caption{Proposed CSI feedback framework CQNet.}
    \label{CQNet}
\end{figure}

We propose an end-to-end ``CQNet" to jointly optimize
CSI dimension reduction, encoding, and CSI reconstruction.
As shown in Fig. \ref{CQNet}, CQNet consists of an encoder DNN
at the UE which includes a dimension compression module and a quantizer,
paired with a decoder network at the gNB. 
The compression module can adopt the encoder neural network of CsiNet or DualNet. 
The quantizer module is parameterized by a trainable 
forward quantization weight vector $\mathbf{w}$ to map 
the unquantized $\mathbf{s}_1$ or $\mathbf{s}_2$, respectively, into an index vector
$\mathbf{k}$ that corresponds to the quantized codeword
$\hat{\mathbf{s}}_1$ or $\hat{\mathbf{s}}_2$. 
An inverse quantization weight vector $\mathbf{v}$ that generates
approximate codeword $\hat{\mathbf{s}}_i$ from the index vector $\mathbf{k}$.  
To facilitate back-propagation during training,
a soft quantizer is used to replace the non-differentiable quantizer function. 
After quantization, the jointly trained decoder DNN at the
receiver is utilized to decode the quantized vector and reconstruct 
the downlink CSI matrix $	\hat{\mathbf{H}}_d$. 


\begin{figure}
    \centering
    \includegraphics[scale=0.45]{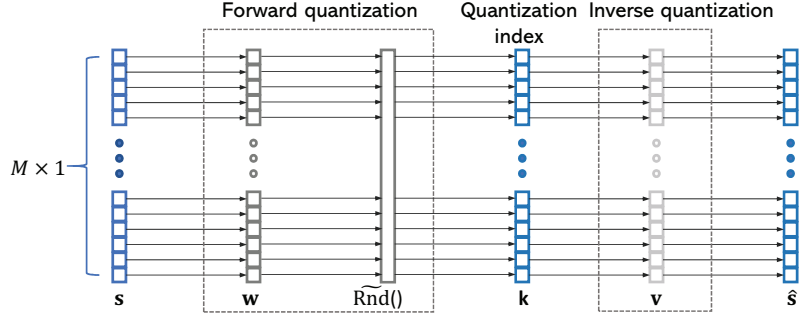}
    \vspace*{-1mm}
    \caption{Illustration of quantizer module. The quantizer has two distinct stages: forward quantization stage and inverse quantization stage. The forward stage maps the input vector $\mathbf{s}_i$ to an integer quantization index vector $\mathbf{k}$ and the reconstruction stage maps the index vector $\mathbf{k}$ back 
 to vector $\hat{\mathbf{s}}_i$ as the approximation of the input vector.}
    \label{Quantizer}
\end{figure}

In this framework, in addition to CSI dimension compression, we also
optimize weight vectors $\{w_i\}$ within the quantization module which
is part of the overall DNN.  We define the weights of the quantizer 
such at the quantization intervals are $d_i = 1/w_i$ for each element
$s_i$ in the compressed CSI vector $\mathbf{s}$. 
As illustrated in
Fig. \ref{Quantizer}, the forward quantization stage can be implemented 
as a set of
element-wise multiplication filters which have only $M$ parameters followed by a 
rounding function. The rounding function can
be viewed as an activation function of a neuron acting on each output from
the element-wise multiplications.
Since the rounding function $\mbox{Rnd}$ has zero gradient almost everywhere, 
slow convergence may take place in back-propagation training.  To overcome this
shortcoming, we propose an approximate
rounding function
\begin{equation}
    \widetilde{\rm Rnd}(x,\ell, r) = \sum_{i=-2^{\ell-1}}^{2^{\ell-1}-1}
    \mbox{sigmoid}[r(x-i-0.5)]-2^{\ell-1},
\end{equation}
which is differentiable and easy
to implement. 
$\widetilde{\rm Rnd}(\cdot)$ is a summation of
sigmoid functions with parameters $\ell$ and $r$ 
where $\ell$ is the number of quantization bits for each 
element in CSI vector $\mathbf{s}$, and $r$ controls the sigmoidal shape. 
After forward quantization stage, an inverse quantization stage is added to reconstruct the approximated codeword $\hat{\mathbf{s}}_i$ using a vector $\mathbf{v}$ before feeding into the decoder module, as shown in Fig. \ref{Quantizer}.

The CQNet framework is compatible with both CsiNet and DualNet-MAG, or
any other CSI compression encoder. 
The CQNet architecture that uses CsiNet for CSI dimension reduction module
and CSI recovery is named CsiQnet. The dimension compression module is followed by the quantizer module shown in Fig. \ref{Quantizer}. The decoder network of CsiNet is 
directly used in CsiQnet to decode the quantized codewords. 
The CQNet architecture that uses DualNet-MAG for CSI dimension reduction module
and CSI recovery is named DualQnet.
DualQnet uses the DualNet-MAG encoder network for dimension compression at the UE, 
and also uses the DualNet-MAG decoder network  at the gNB 
to recover the CSI. Quantizer module is inserted between the dimension compression module and decoder module.

We define the loss function of CsiQnet or DualQnet:
\begin{equation}
    L(\hat{\mathbf{H}}_d, \mathbf{w}) = L_m(\hat{\mathbf{H}}_d,\mathbf{H}_d) + \lambda L_{quan}(\mathbf{w}),
    \label{jointcost}
\end{equation}
as a combination of mean square error (MSE)
loss $L_m$ and a quantization loss $L_{\rm quan}$. 
$L_{quan}$ is used for the regularization that 
accounts for quantization efficiency and convergency. 
One simple function for this purpose is $L_{quan}(\mathbf{w}) = \left\| \mathbf{w} \right\|$. The training objective
is to find the encoding and decoding parameters which can achieve the optimum CSI reconstruction accuracy given the specific quantization bits per value $\ell$.

Through training based on a large MIMO CSI data set generated using
well known practical channel models such as  COST 2100 model \cite{c2100},  
CsiQnet and DualQnet can converge to optimized settings.  During live downlink CSI
feedback, both CsiQnet and DualQnet can generate more efficiently quantized codewords 
$\hat{\mathbf{s}}_i$ which can significantly improve the
accuracy of CSI reconstruction at fixed bandwidth or bitwidth. 
CsiQnet and DualQnet enable 
more effective bandwidth usage with little CSI reconstruction loss.

To train the model, normalization is applied 
in both downlink and uplink CSI matrices.
Adam optimizer is adopted to update the 
DL network parameters. 
The initial learning rate is set to $0.001$. 
To accelerate the convergence speed of the training, 
we utilize the weights trained in CsiNet and DualNet-MAG to initialize the dimension compression modules and decoder networks of CsiQnet and DualQnet, respectively.
Notice that DualQnet optimizes the magnitude feedback during joint training 
to minimize (\ref{jointcost}).  For the separate phase feedback of
compressed CSI, 
we shall design another MAPQ DNN in Section \ref{sectionMAQ}. 


\subsection{DL-based Phase Quantization }
\label{sectionMAQ}

DualNet-MAG utilizes the magnitude correlation of bi-directional CSIs
in polar coordinate to reduce the amount of feedback for CSI magnitudes 
and improve the CSI feedback efficiency.
However, weak phase correlation between uplink/downlink CSI requires
the UE to efficiently quantize and encode all downlink CSI phases
for feedback. 
However, it is well known that uniform phase quantization results in
unnecessarily fine quantization at low magnitude and coarse quantization 
at high magnitude.  
Therefore, bandwidth efficiency phase quantization is an important
issue to tackle in DualQnet.

Let $\mathbf{M} = [\mathbf{M}_{i,j}] $ be the magnitudes of CSI matrix, $\mathbf{P} = [\mathbf{P}_{i,j} ]$ be
the phases of CSI matrix, $\hat{\mathbf{M}}$ be the recovered magnitudes of CSI matrix, and $\hat{\mathbf{P}}$ be the recovered phases of CSI matrix, respectively.
We can write a matrix $e^{j {\mathbf{P}}}$ whose elements are $e^{\mathbf{P}_{i,j}}$. 
The optimization objective of DualNet-MAG is to minimize the MSE of recovered CSI matrices, i.e., $\mathbf{E} \{\left \|  \mathbf{H}_d - \hat{\mathbf{H}}_d\right \|^2\} = 
\mathbf{E}\{\left \|\mathbf{M}\odot e^{j {\mathbf{P}}} - \hat{\mathbf{M}}
\odot
e^{j\hat{\mathbf{P}} }\right\|^2\}$, where $\odot$ denotes Hadamard matrix product. 
It is challenging to solve this problem since the recovered magnitude and phase influence the MSE jointly.  To reduce the complexity of this problem, we 
consider the upper bound:
\begin{equation}
\begin{aligned}
& \left \|\mathbf{M}\odot e^{j {\mathbf{P}}} - \hat{\mathbf{M}}\odot e^{j\hat{\mathbf{P}} }\right \|^2 \\ 
&= \left \|\mathbf{M}\odot e^{j {\mathbf{P}}} - \mathbf{M}\odot e^{j {\hat{\mathbf{P}}}}+\mathbf{M}\odot e^{j {\hat{\mathbf{P}}}}- \hat{\mathbf{M}}\odot 
e^{j\hat{\mathbf{P}} }\right \|^2 \\
& \leqslant 2 \left \|\mathbf{M}\odot 
e^{j {\mathbf{P}}} - \mathbf{M}\odot e^{j {\hat{\mathbf{P}}}}\right \|^2+2\left \|\mathbf{M}\odot e^{j {\hat{\mathbf{P}}}}- \hat{\mathbf{M}}\odot 
e^{j\hat{\mathbf{P}} }\right \|^2 \\
&= 2 \left \|\mathbf{M}\odot (e^{j {\mathbf{P}}} 
- e^{j {\hat{\mathbf{P}}}})\right \|^2 + 2\left \|\mathbf{M}- \hat{\mathbf{M} }\right \|^2.
    \end{aligned}
\end{equation}
Consequently, the optimization goal can be relaxed to minimize 
\[ \mathbf{E}_{(\mathbf{M},\mathbf{P})} \left (\left \|\mathbf{M}\odot (e^{j {\mathbf{P}}} - e^{j {\hat{\mathbf{P}}}})\right \|^2 \right ) + \mathbf{E}_{\mathbf{M}} \left ( \left \|\mathbf{M}- \hat{\mathbf{M} }\right \|^2 \right).
\]
DualNet can minimize $\mathbf{E}_{\mathbf{M}}  ( \|\mathbf{M}- \hat{\mathbf{M} } \|^2)$. It is clear that the first part $\mathbf{E}_{(\mathbf{M},\mathbf{P})} (\|\mathbf{M}\odot (e^{j {\mathbf{P}}} - e^{j {\hat{\mathbf{P}}}}) \|^2 )$ represents
phase quantization error amplified by the corresponding magnitude.

Therefore, to further reduce feedback bandwidth, CSI quantization error 
can be kept small by applying the magnitude-adaptive phase quantization (MAPQ) 
principle in which CSI coefficients with larger magnitude
adopt finer phase quantization, and vice versa. 
After recovering the magnitude, gNB can restore the quantified phase based on MAPQ.  
Such MAPQ can keep the quantization error close for a range of
magnitudes. 
Since quantization bits of phase vary with the magnitude, 
the expectation of quantization bits depends on the 
distribution of 
CSI magnitude. 
Thus, we need to allocate the the number of phase quantization bits 
based on the distribution of CSI magnitude underlimited average bitwidth.
Such problems typically become a mixed integer nonlinear programming problem 
as described in \cite{nlp1}, which is NP-hard.

A heuristic quantization bit
allocation solution was provided in \cite{dlo} based on examining the
data set. In this method,  the cumulative distribution function (CDF) 
of CSI magnitudes is estimated to determine magnitude values corresponding to  
CDF value of 0.5, 0.7, 0.8, and 0.9, respectively. 
These four points divide the CSI magnitudes into five ordered 
segments from low to high. Accordingly,  3, 4, 5, 6, 7 
phase quantization bits are allocated, respectively, to encode
the CSI phases. 
This set of MAPQ codewords given in \cite{dlo} can
generate codeword of the mean length of 4.1 bits to
achieve the same MSE as that obtained using a 6-bit uniform quantizer.

However, the heuristic method of \cite{dlo} is inflexible 
with respect to the segments. Given different mean bitwidth
constraints, we have to determine different allocation based on
heuristics. Thus, we propose a more flexible and general
design method in this work. 

\begin{figure}
    \centering
    \includegraphics[scale=0.5]{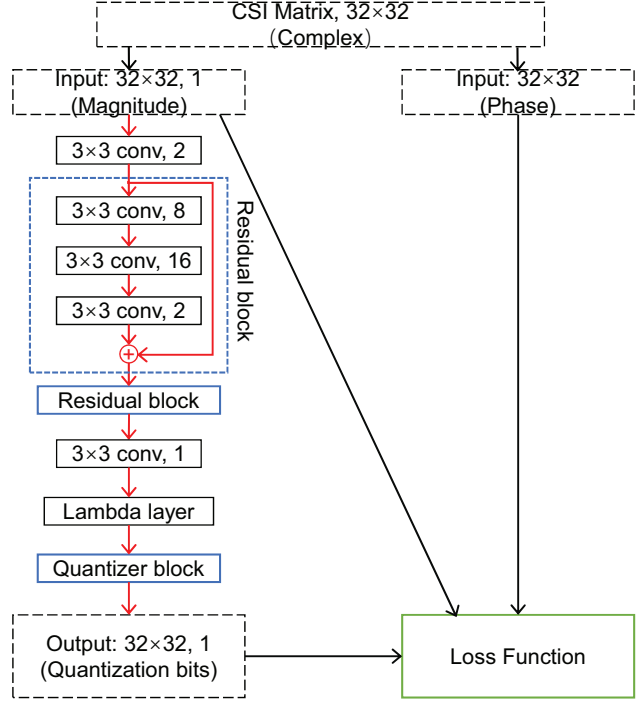}\vspace*{-1mm}
    \caption{Illustration of magnitude-adaptive phase quantization.}
    \label{PhaseQuan}
\end{figure}

We propose an innovative DNN to solve the bit allocation problem. 
This new framework PhaseQuan can optimize the allocation of MAPQ
quantization bits for phase based on unsupervised learning. As shown in Fig. \ref{PhaseQuan},  
PhaseQuan utilizes the magnitudes as the input to DL network and 
includes CSI phases, magnitudes and corresponding quantization bits 
in its loss function. The input of magnitudes pass through
a  $3 \times 3$ convolutional layer and two residual block, 
which explore the potential spatial correlation within the CSI matrix.
The ensuing convolutional layer, lambda layer,
and quantizer block are used to project the output of the
last hidden layer to quantization bits. 
The convolutional layer first utilizes the 
``sigmoid" activation to project the input within $[0,1]$. 
The lambda layer is defined as $log_2(\frac{1}{x+\epsilon})$, 
in which $\epsilon>0$ is a small value to ensure non-zero denominator.
Here, 
 $\frac{1}{x+\epsilon}$ corresponds to the number of quantization intervals. 
Logarithm $log_2(\cdot)$ and the quantizer module can map the number of 
quantization intervals to the number of quantization bits. 

To minimize $\mathbf{E}_{(\mathbf{M},\mathbf{P})}  ( \|\mathbf{M} \odot(e^{j {\mathbf{P}}} - e^{j {\hat{\mathbf{P}}}}) \|^2  )$ within the constraint on 
the quantization bitwidth,  
we adopt the entropy of phase as the optimization
regularizer as more quantization bits 
lead to higher phase entropy. Thus, we propose a loss function 
\begin{equation}
    L(\mathbf{M}, \mathbf{P}, \mathbf{Y}) = L_m(\mathbf{M},\mathbf{P},\mathbf{Y}) + \lambda L_{y}(\mathbf{Y}),
\end{equation}
where $\mathbf{Y}$ is a matrix of integer elements
representing the number of quantization bits for the corresponding
CSI matrix element, optimized by the PhaseQuan.  Since the quantization bits for each phase is variable, the codes are not uniquely decodable without knowing the quantization bits for each phase. By approximating $\mathbf{M}$ using $\hat{\mathbf{M}}$, the gNB can infer the corresponding bits of each phase from the recovered magnitude for uniquely decoding. Consequently, the phase quantization error evaluation function is set to be 
\[ L_m(\mathbf{M},\mathbf{P},\mathbf{Y}) = \mathbf{E}_{(\mathbf{M},\mathbf{P})}\left (\left \|\hat{\mathbf{M}}\odot e^{j(\hat{\mathbf{P}} - {\mathbf{P}})}\right \|^2\right ),
\]
in which $
\hat{\mathbf{P}} $
is the quantized phase matrix. Additionally, define an entropy
$H(\check{\mathbf{P}}_{i,j})$ for quantized phase
$\check{\mathbf{P}}_{i,j}$ that corresponds to 
magnitude $\mathbf{M}_{i,j}$.
In this paper, we assume the phase to be uniformly distributed
over $2\pi$ \cite{phase_distribution}. 
Consequently, $H(\check{\mathbf{P}}_{i,j}) = \mathbf{Y}_{i,j}$.
As a result, we use
\[ L_y(\mathbf{Y}) = \mathbf{E}_{(\mathbf{M},\mathbf{P})}
\left (\frac{1}{Q_f \times N_b}\sum_{i,j} H(\check{\mathbf{P}}_{i,j})\right )
\] 
as an entropy regularizer to reduce number of quantization bits in phase feedback.  
Adjustable parameter $\lambda$ value governs
the trade-off between the quantization bits and the reconstruction loss. 


\section{Performance Evaluation}

\subsection{Experiment Setup}
We use the industry grade COST 2100 model \cite{c2100}
to generate massive MIMO channels for both training and testing of 
our DNN architecture. The training sample size is $70,000$ and testing sample size is $30,000$. 
The values of epoch and batch size are set to $600$ and $200$, respectively.
We test two 
scenarios:
\begin{itemize}
	\item[(a)] indoor channels with 5.1\,GHz uplink band and 5.3\,GHz
	downlink center frequency.  
	\item[(b)] semi-urban outdoor channels with 850~MHz uplink band
	and 930~MHz downlink center frequency. 
	\end{itemize}
Uplink and downlink bandwidths of 20 MHz and 5 MHz are selected for the indoor and outdoor scenarios, respectively.

We place gNB at the center of a square area of lengths $20$m for indoor coverage and $400$m for outdoor
coverage, respectively. 
We randomly position UEs within the coverage area. The
gNB uses ULA with $N_b = 32$ antennas and $N_f = 1024$ subcarriers. 
After transforming the channel matrix $\mathbf{H}_f$ into the delay domain
$\mathbf{H}_t$, only the first 32 rows are kept for feedback 
reporting due to sparsity. 

To evaluate the accuracy of CSI recovery, we use normalized MSE 
\begin{equation}
\textrm{NMSE} = \frac{1}{n}\sum_{k=1}^n\Arrowvert\mathbf{H}_d^k-\mathbf{\hat{H}}_d^k\Arrowvert^2/\Arrowvert\mathbf{H}_d^k\Arrowvert^2,
\end{equation}
where $k$ and $n$ are the index 
and total number of samples in the testing set, 
respectively. We compare the CsiQnet and DualQnet with the CsiNet and DualNet-MAG 
for two quantization methods, respectively. 
The uniform quantization (UQ) and $\mu$-law non-uniform quantization are adopted ($\mu Q$). For the  $\mu$-law quantization, we use $\mu=255$ in the companding function
\begin{equation}
    F(x)=\mbox{sgn}(x)\frac{\ln(1+\mu \left | x \right |)}{\ln(1+\mu)}.
\end{equation}

For the CsiQnet and CsiNet, CSI matrix is divided into 
two (real and imaginary) channels as the input to the
DNN.  We compare the CSI reconstruction performance 
under the compressed dimension $M = 64$ and $256$. 
For DualQnet and DualNet-MAG,  
we compare the CSI reconstruction performance under the compressed 
dimension $M = 32$ and $128$.

\subsection{CSI Reconstruction Performance Evaluation}

\begin{figure}  
	 \hspace*{-3mm}
\subfigure[CsiQnet indoor] {\label{figpe:a} 
\includegraphics[width=0.51\columnwidth]{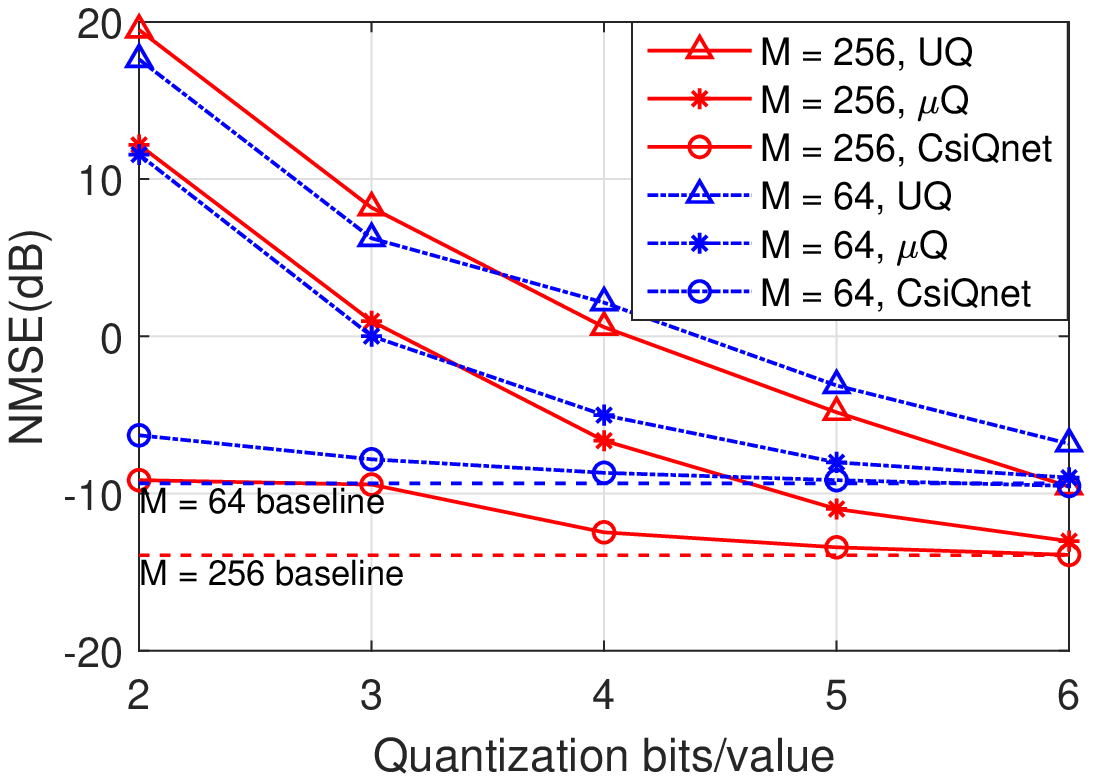}
} \hspace*{-8mm}
\subfigure[DualQnet indoor] { \label{figpe:b} 
\includegraphics[width=0.51\columnwidth]{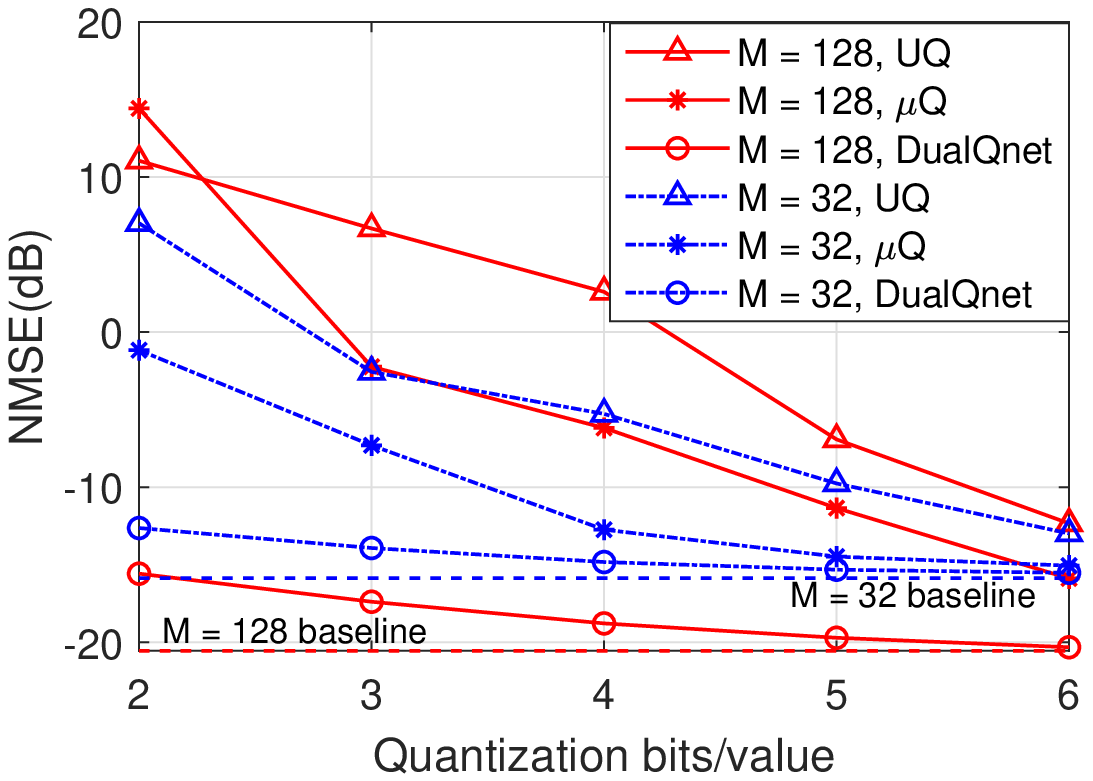} 
}   \hspace*{-3mm}
\subfigure[CsiQnet outdoor] {\label{figpe:c} 
\includegraphics[width=0.51\columnwidth]{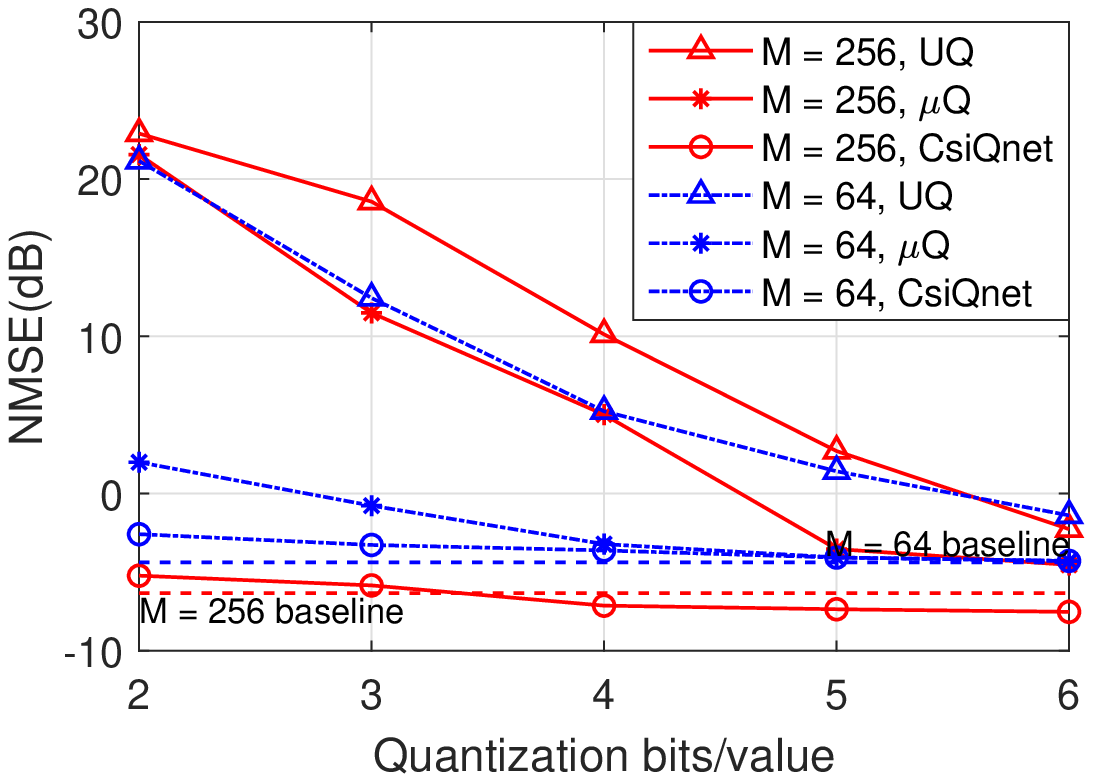}
}   \hspace*{-8mm}
\subfigure[DualQnet outdoor] { \label{figpe:d} 
\includegraphics[width=0.51\columnwidth]{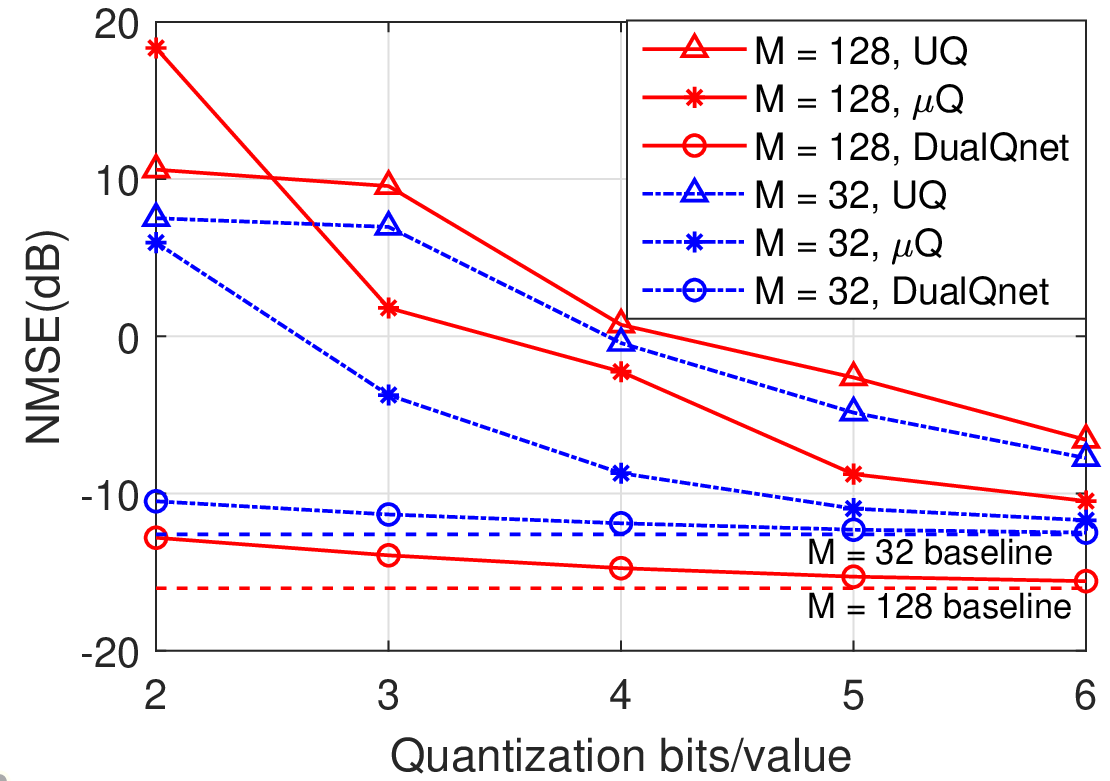} 
} 
\caption{CSI recovery comparison at various quantization levels.} 
\label{figurepe} \vspace*{-3mm}
\end{figure}

We compare the CSI reconstruction accuracy achieved from 
the above DL networks under five different bitwidths
of 2,3,4,5 and 6. Single precision is used as the baseline 
for performance comparison.  Fig. \ref{figurepe} shows 
the NMSE performance of our proposed
CsiQnet and DualQnet.

As shown in Fig. \ref{figurepe},  CsiQnet and DualQnet outperform CsiNet and DualNet-MAG using UQ and $\mu$Q obviously, and can achieve comparable performance with single precision data type using only $5$ bits. As expected, non-uniform $\mu$-law quantization ($\mu$Q) outperforms uniform quantization (UQ) obviously, except for the extreme cases where the NMSE is greater than 10 dB. For the CsiQnet, 2 bits quantization can achieve comparable performance as the $\mu$Q using 5 bits in indoor case and 6 bits in outdoor case. For DualQnet, 2 bits quantization can achieve even
better performance than 6-bit $\mu$Q.

From  the  results  in  Fig. \ref{figurepe}, 
we observe more robust performance when compressed dimension $M$ is relatively
small. The NMSE degrades faster when $M$ is large. 
The possible reason is that the lower dimension compression 
relies on principal components, 
while high-accuracy reconstruction further requires more
detailed information of compressed vectors. Thus, smaller feedback
error can lead to a larger degradation in reconstruction accuracy 
when compression ratio is low. 
Interestingly, CsiQnet could outperform the baseline in outdoor scenario for $M = 256$. One possible reason is that the 
quantized value can help DNN overcome
some cases of local minima during training. 

\begin{figure} \hspace*{-3mm}
\subfigure[CsiQnet indoor] {\label{figpe2:a} 
\includegraphics[width=0.52\columnwidth]{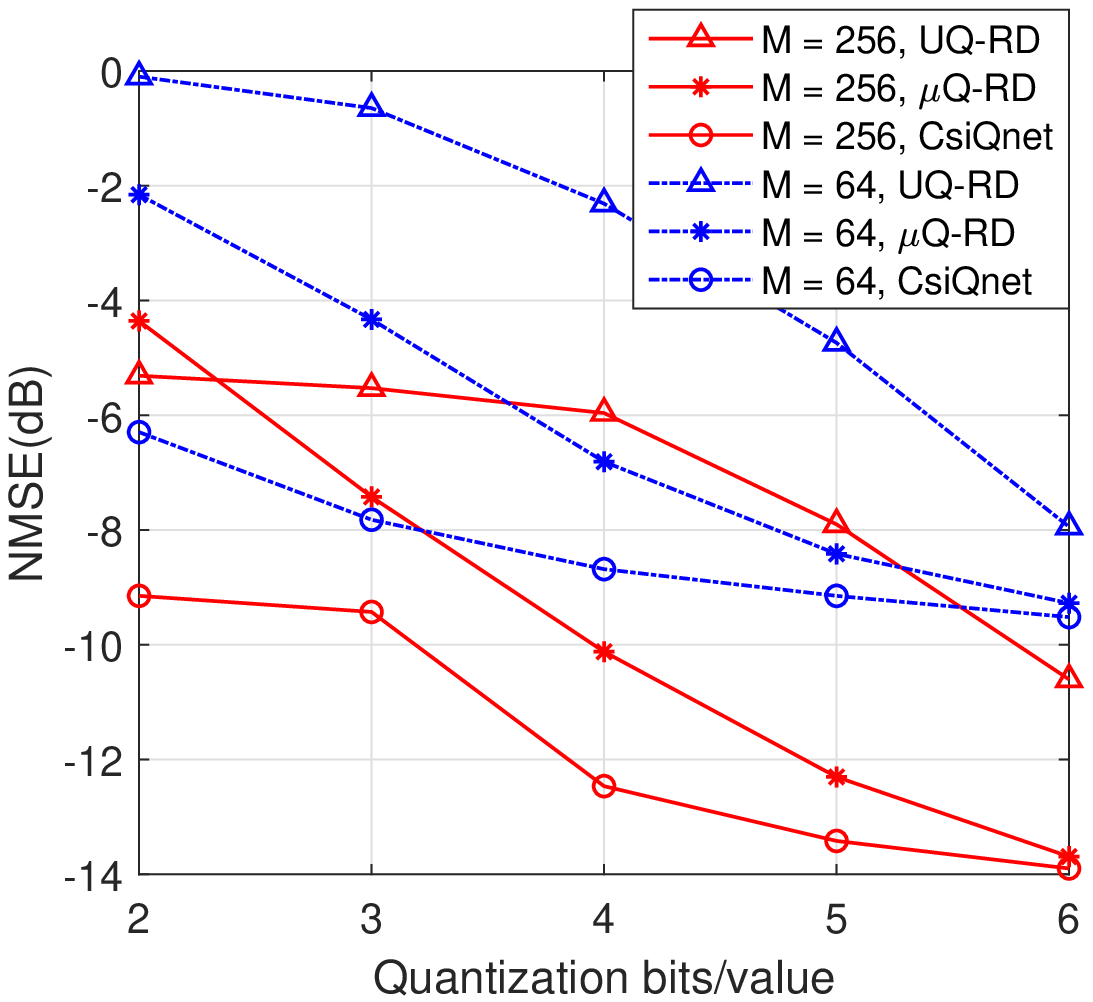}
} \hspace*{-8mm}
\subfigure[DualQnet indoor] { \label{figpe2:b} 
\includegraphics[width=0.52\columnwidth]{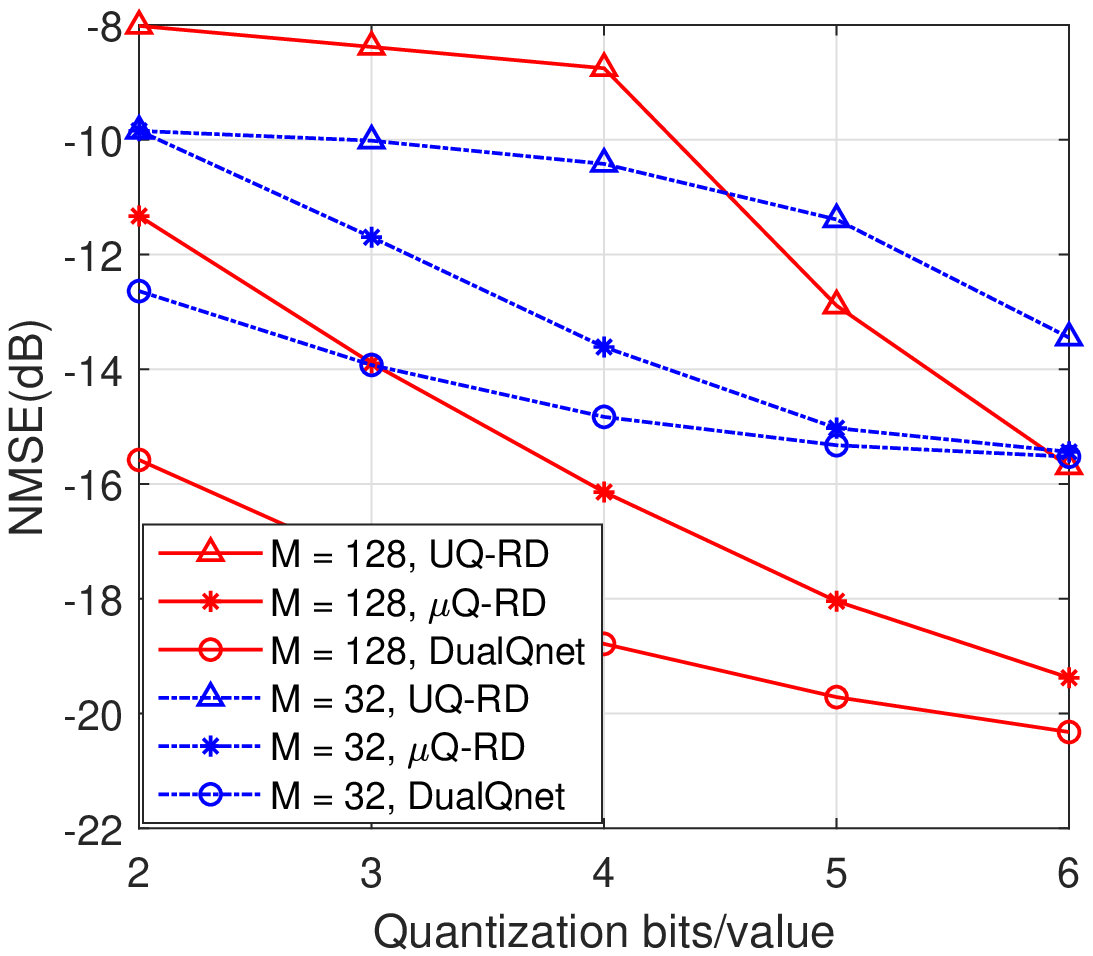} 
} \hspace*{-3mm}
\subfigure[CsiQnet outdoor] {\label{figpe2:c} 
\includegraphics[width=0.52\columnwidth]{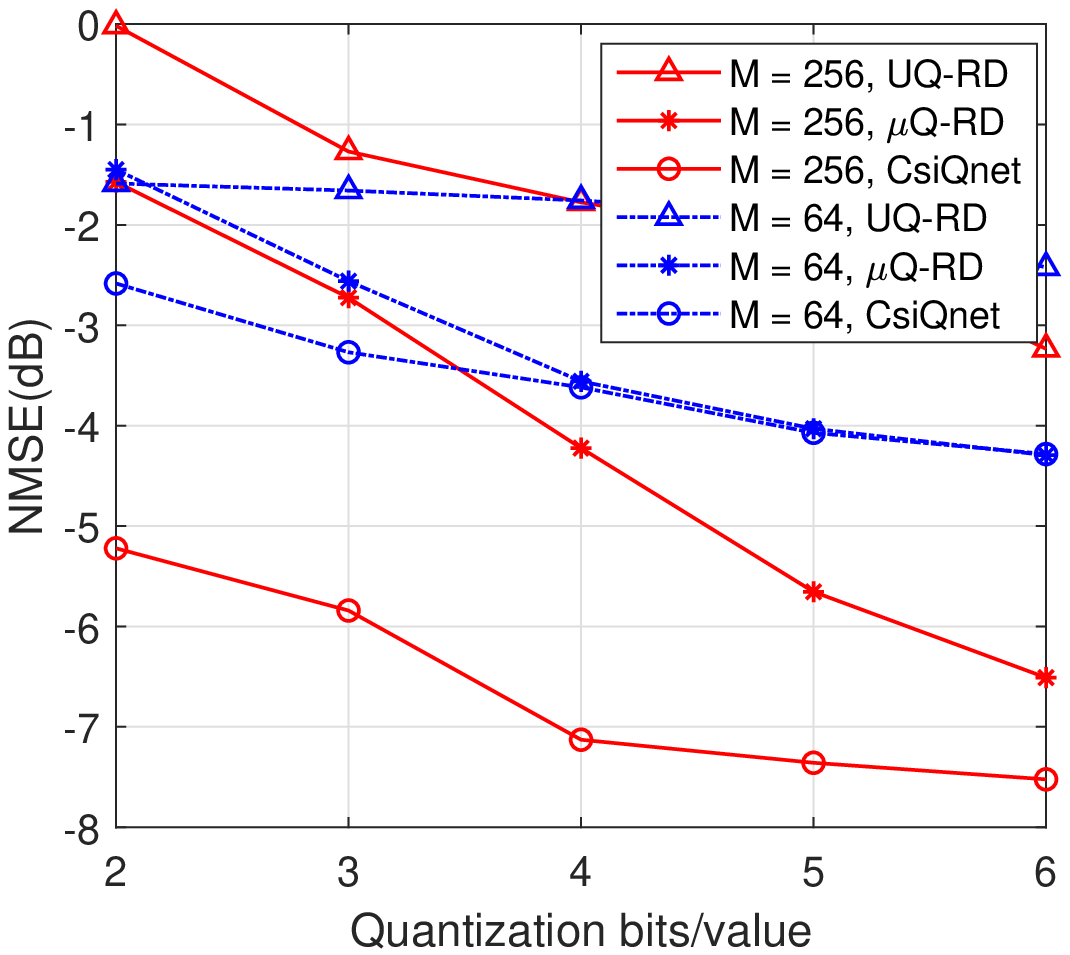}
} \hspace*{-8mm}
\subfigure[DualQnet outdoor] { \label{figpe2:d} 
\includegraphics[width=0.52\columnwidth]{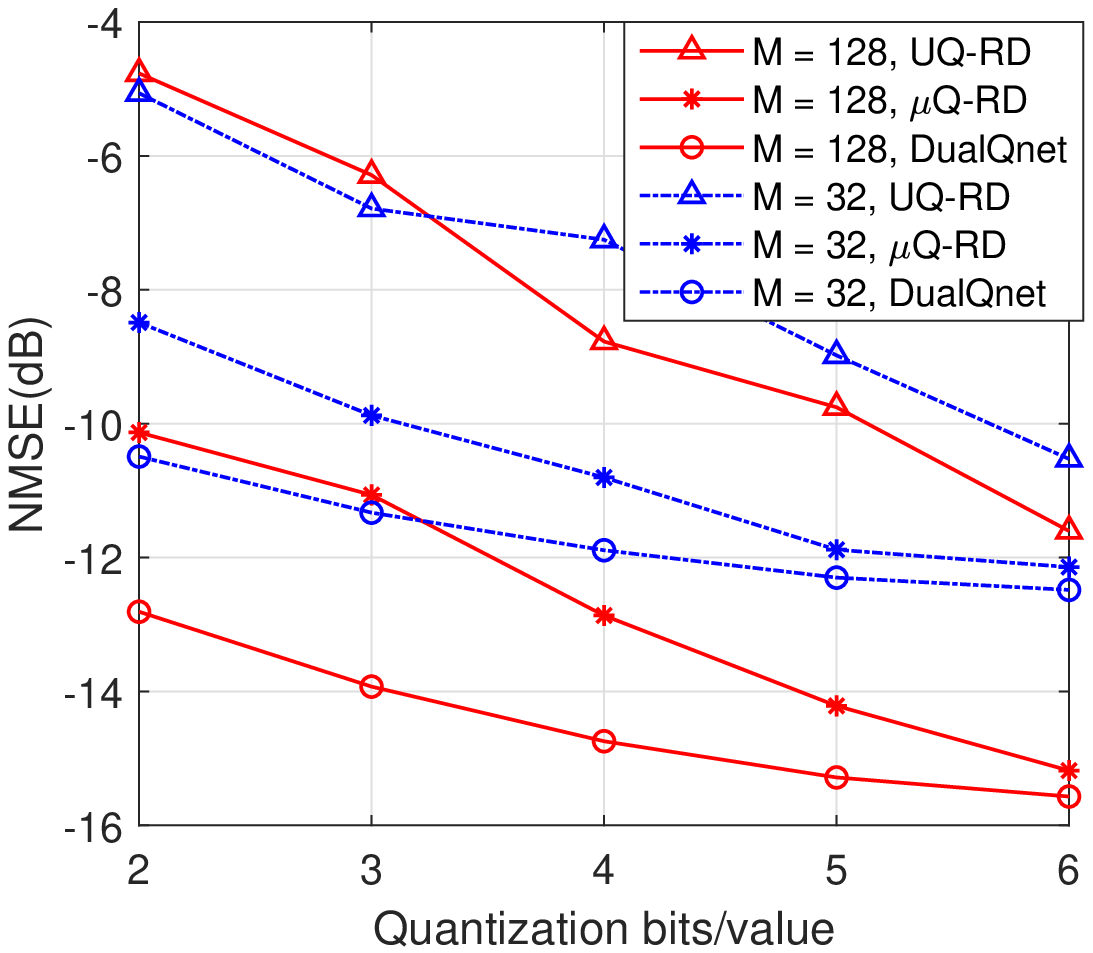} 
} 
\caption{CSI recovery comparison at different quantization levels between CQNet and retrained decoder networks (RD).} 
\label{figurepe2} 
\end{figure}

For a more comprehensive comparison, we 
further retrain the decoder networks of CsiNet and 
DualNet-MAG after quantizing the codewords from their 
encoder networks, 
and compare their NMSE with CsiQnet and DualQnet. Both
UQ and $\mu $Q are used to quantize the codewords 
from CsiNet and DualNet-MAG encoders. 
Denote the retrained decoder network results by RD,
Fig. \ref{figurepe2} shows the NMSE performance comparison of our 
proposed CsiQnet and DualQnet with CsiNet and DualNet-MAG 
when used with the retrained decoder. 
As shown in Fig. \ref{figurepe2},  CsiQnet and DualQnet both 
outperform CsiNet and DualNet-MAG under quantization and the retrained decoder.
In fact, the accuracy gap increases with decreasing  
number of quantization bits/value.  
This result demonstrates that our end-to-end DNN framework jointly integrating
CSI compression, quantization with reconstruction 
achieves better performance than combining individually optimized modules. 
Fig. \ref{figurepe2} also shows that  $\mu $Q  
generally delivers a better performance than UQ.


\begin{table}[]
\centering
\caption{Average required bits/value after entropy encoding when the quantization bits/value are 5 and 6.}
\label{entropy enc} 
\begin{tabular}{c|c|c|c|c|c|c}
\hline
Scenario                 & Network                 & Dimension & \begin{tabular}[c]{@{}c@{}}$\mu$Q\\ 5 bits/value\end{tabular} & \begin{tabular}[c]{@{}c@{}}CQNet\\ 5 bits/value\end{tabular} & \begin{tabular}[c]{@{}c@{}}$\mu$Q\\ 6 bits/value\end{tabular} & \begin{tabular}[c]{@{}c@{}}CQNet\\ 6 bits/value\end{tabular} \\ \hline
\multirow{4}{*}{Indoor}  & \multirow{2}{*}{CsiNet}      & 64        & 4.36                                                    & 3.88                                                   & 5.37                                                    & 4.41                                                   \\ \cline{3-7} 
                         &                              & 256       & 4.32                                                    & 3.91                                                   & 5.34                                                    & 4.46                                                   \\ \cline{2-7} 
                         & \multirow{2}{*}{DualNet} & 32        & 4.38                                                    & 4.28                                                   & 5.39                                                    & 4.54                                                   \\ \cline{3-7} 
                         &                              & 128       & 4.37                                                    & 3.71                                                   & 5.38                                                    & 4.15                                                   \\ \hline
\multirow{4}{*}{Outdoor} & \multirow{2}{*}{CsiNet}      & 64        & 4.15                                                    & 3.04                                                   & 5.17                                                    & 3.58                                                   \\ \cline{3-7} 
                         &                              & 256       & 4.30                                                    & 3.75                                                   & 5.31                                                    & 4.47                                                   \\ \cline{2-7} 
                         & \multirow{2}{*}{DualNet} & 32        & 4.28                                                    & 4.23                                                   & 5.30                                                    & 4.96                                                   \\ \cline{3-7} 
                         &                              & 128       & 4.27                                                    & 3.88                                                   & 5.28                                                    & 4.43                                                   \\ \hline
\end{tabular}
\end{table}

Once the CSI feedback is quantized, we can exploit entropy 
encoding to further compress the CSI feedback.
Entropy encoding is a lossless scheme to 
compress digital data. We select the arithmetic coding \cite{arithmetic_encoding}, 
which is a simple and common entropy encoding to encode the quantized CSI 
coefficients. We consider $\mu$Q and CQNet with 5 and 6 bits per CSI value
as examples.  The average required numbers of bits per CSI value
after entropy encoding are given in Table \ref{entropy enc}. As shown in Table \ref{entropy enc}, entropy encoding can help
CQNet save additional 1 and 1.5 bits per CSI value on average 
when the quantization bits/value are 5 and 6, respectively. 
It means that only 4 bits per CSI value
are required by entropy encoding to deliver comparable performance 
previously achieved by float32 data type. We also find that,
compared with $\mu$Q, CQNet saves more bitwidth and achieves higher CSI 
reconstruction accuracy. In other words, 
although the entropy of CQNet is lower than that of $\mu$Q, 
CQNet packs more useful information in its codewords for CSI reconstruction.

\subsection{Robustness Evaluation}

\begin{figure} [t]
	\hspace*{-3mm}
\subfigure[CsiQnet indoor] {\label{figpe3:a} 
\includegraphics[width=0.52\columnwidth]{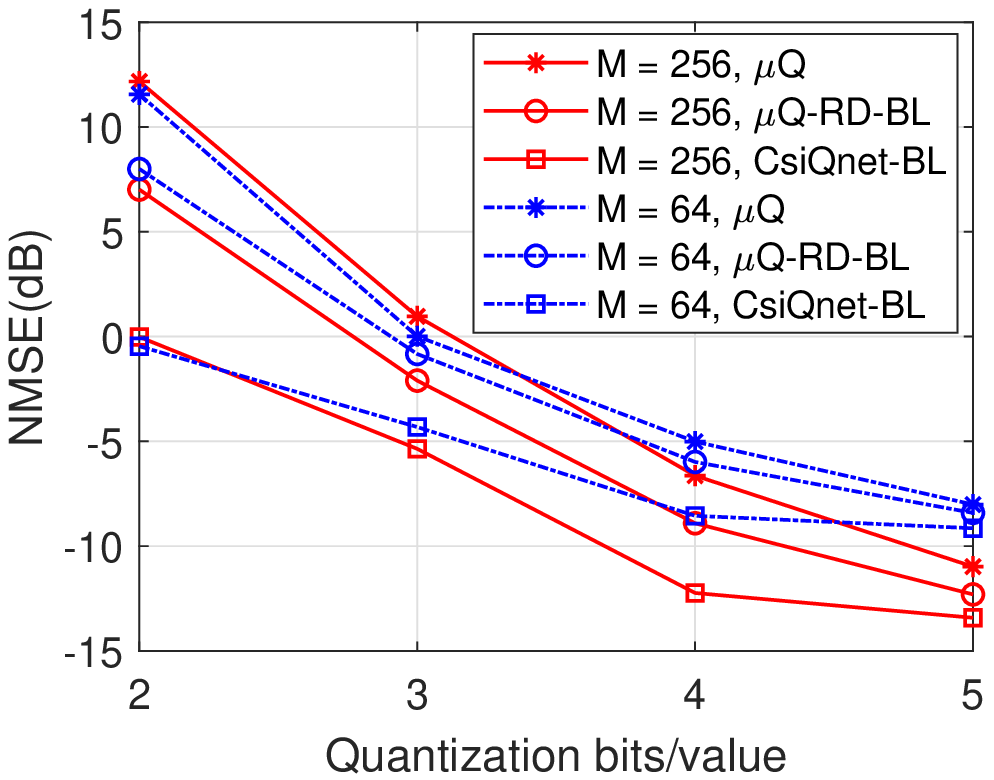}
} \hspace*{-8mm}
\subfigure[DualQnet indoor] { \label{figpe3:b} 
\includegraphics[width=0.52\columnwidth]{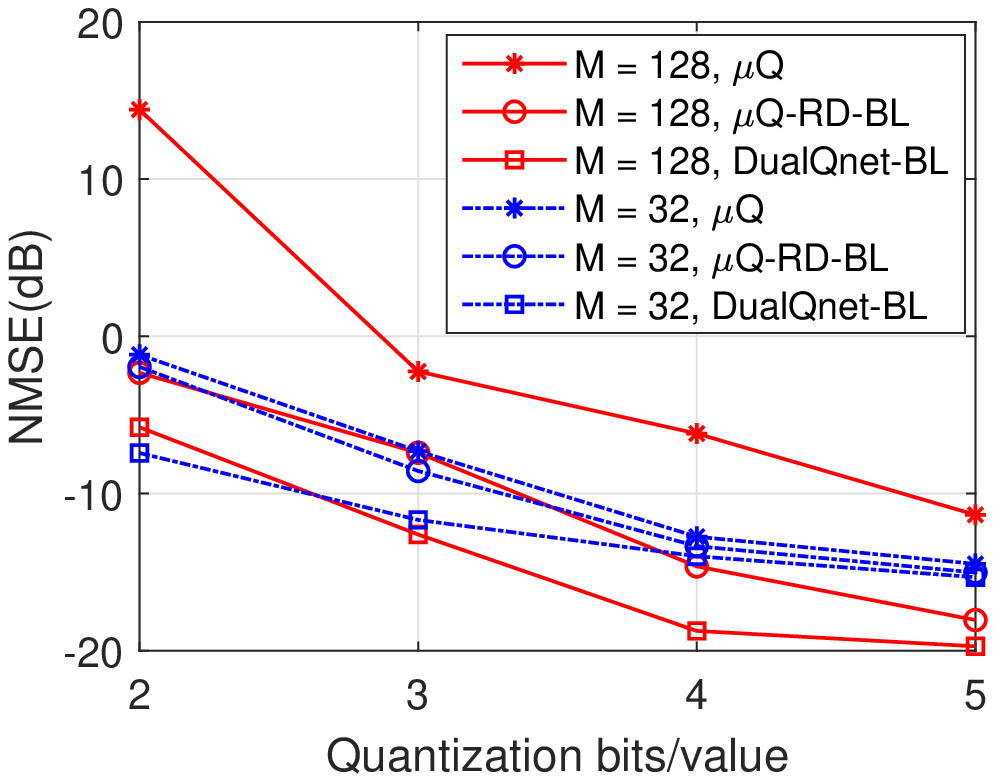} 
} \hspace*{-3mm}
\subfigure[CsiQnet outdoor] {\label{figpe3:c} 
\includegraphics[width=0.52\columnwidth]{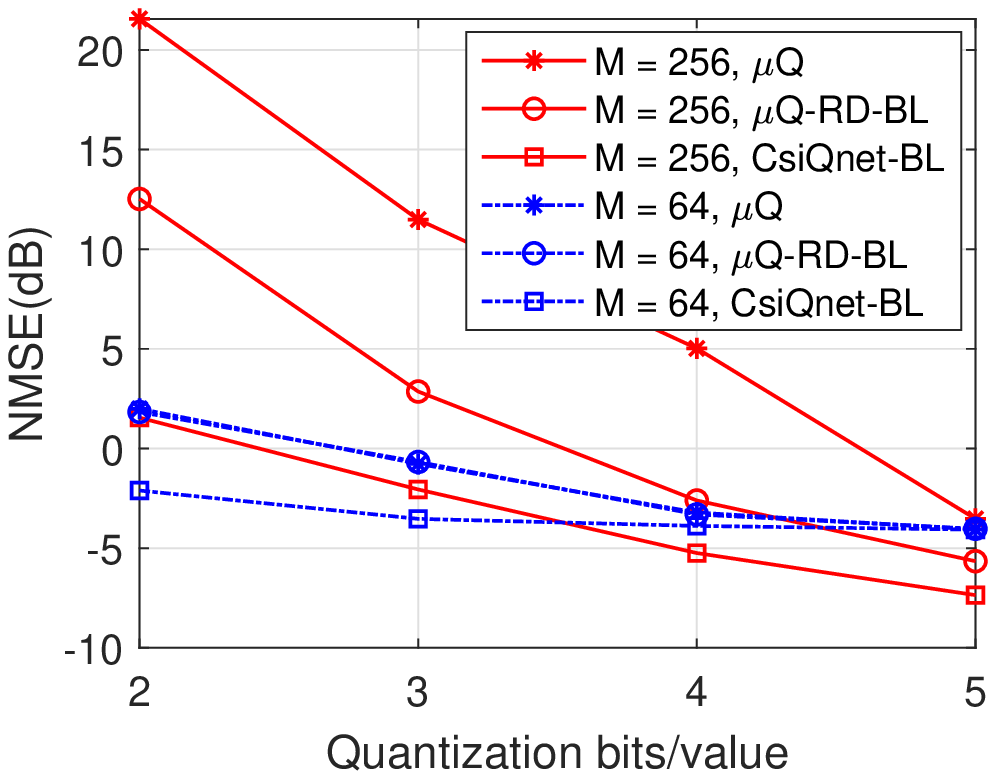}
} \hspace*{-8mm}
\subfigure[DualQnet outdoor] { \label{figpe3:d} 
\includegraphics[width=0.52\columnwidth]{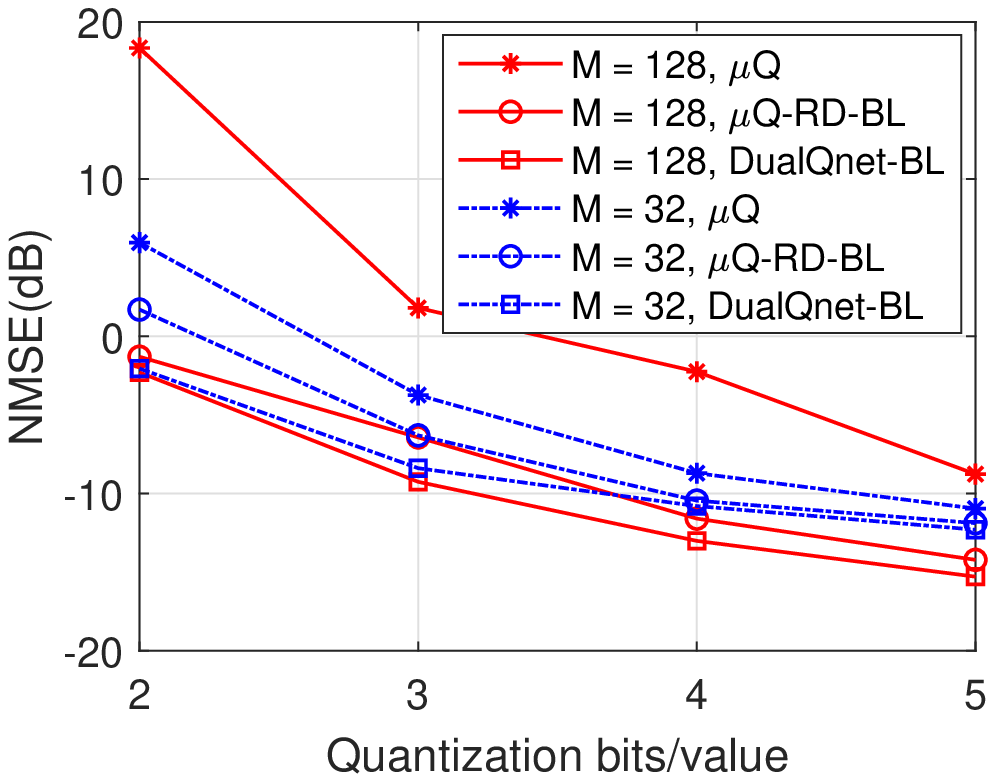} 
} 
\caption{CSI recovery comparison at different quantization levels considering band limitation.} 
\label{figurepe3}
\end{figure}
To further test the robustness of CsiQnet and DualQnet, we consider the case
when there is only one neural network trained for the codewords quantization and reconstruction but different bitwidths may be 
required due to the bandwidth changes. 
Fig. \ref{figurepe} shows that 
CsiQnet and DualQnet using 5 bits can achieve the similar performance as CsiNet and DualNet-MAG using single precision. As a result, we can 
select 5 quantization bits/value as an example to evaluate the robustness of CsiQnet and DualQnet. Since $\mu $Q outperforms UQ clearly when quantization bits/value is 5, 
we select $\mu $Q and $\mu $Q-RD for CsiNet and DualNet-MAG as a 
reference to test the robustness.

Considering a DNN already trained for the 5 quantization bits/value.
If the UE does not have sufficient bandwidth to transmit each codeword using 5 bits,
we need to further shorten the codewords using fewer bits: 2, 3 and 4 bits/value. Fig. \ref{figurepe3} shows the performance of CsiQnet and DualQnet
under different bandwidth limitations (BL). As shown in Fig. \ref{figurepe3}, the performance of CsiQnet and DualQnet in BL cases using fewer bits for the quantized codewords clearly outperforms the CsiNet and DualNet-MAG using $\mu$Q and $\mu $Q-RD. 
On the other hand, although CsiQnet and DualQnet in BL cases are more robust 
than other methods, their performance has a relatively obvious degradation when the quantization bits/value falls below 4. 
In our future work, we plan to extend the CsiQnet and DualQnet to 
variable quantization bitwidth.


\subsection{Quantization Evaluation}
In this subsection, we analyze the performance of our quantizer module, and try to go inside the DL networks to explore why CsiQnet and DualQnet outperform the UQ and $\mu $Q methods.

We first compare the performance of our approximated rounding
function with the actual rounding function, and show the MSE between the output of $\widetilde{\rm Rnd}(\cdot)$ and round function in Numpy. Table \ref{MSE} 
shows the MSE under different number of quantizations bits. 
As shown in Table \ref{MSE}, MSE generally is near $-24$ dB, which is negligible compared with the rounded integer. We also compare the NMSE of the CSI reconstruction between using $\widetilde{\rm Rnd}(\cdot)$ and ${\rm Rnd}(\cdot)$.
The differences are less than $-0.15$ dB,
which means the approximate 
$\widetilde{\rm Rnd}(\cdot)$ works well for the quantization.
\begin{table}
\centering
	\caption{MSE(dB) of the approximated rounding function}
	\label{MSE}
	\begin{tabular}{c|c|c|c|c|c}
		\hline
		&      & \multicolumn{2}{c|}{Indoor} & \multicolumn{2}{c}{Outdoor} \\ \hline
		\multirow{6}{*}{CsiQnet}  & Bits & $M$ = 64       & $M$ = 256      & $M$ = 64        & $M$ = 256      \\ \cline{2-6} 
		& 2    & -24.4319     & -24.4578     & -21.7272      & -24.4584     \\ \cline{2-6} 
		& 3    & -24.2357     & -24.1985     & -23.9548      & -24.2606     \\ \cline{2-6} 
		& 4    & -24.1401     & -24.1621     & -24.0639      & -24.1691     \\ \cline{2-6} 
		& 5    & -24.1564     & -24.1512     & -24.1445      & -24.1482     \\ \cline{2-6} 
		& 6    & -24.0967     & -24.1336     & -24.1289      & -24.1554     \\ \hline
		\multirow{6}{*}{DualQnet} & Bits & $M$ = 32       & $M$ = 128      & $M$ = 32        & $M$ = 128      \\ \cline{2-6} 
		& 2    & -24.3763     & -24.9159     & -24.1652      & -24.4255     \\ \cline{2-6} 
		& 3    & -24.2162     & -24.2249     & -24.1463      & -24.2362     \\ \cline{2-6} 
		& 4    & -24.2008     & -24.1677     & -24.1937      & -24.1402     \\ \cline{2-6} 
		& 5    & -24.1409     & -24.1249     & -24.2236      & -24.1325     \\ \cline{2-6} 
		& 6    & -24.1286     & -24.1522     & -24.0995      & -24.131      \\ \hline
	\end{tabular}
\end{table}

It would be helpful for us to examine the effect of quantization error
empirically to understand the performance of different CSI feedback
methods under study. 

We measure the normalized mean square quantization error (NMSQE) 
defined as $\mathbf{E}[\frac{\Arrowvert\mathbf{s}-\hat{\mathbf{s}}\Arrowvert^2}{\Arrowvert\mathbf{s}\Arrowvert^2}]$.  
We choose 5 quantization bits.
We select CsiNet with $M = 256$ and DualNet-MAG with $M = 128$
to demonstrate the quantization error of UQ and $\mu$Q.
As shown in Table \ref{NMSE},  NMSQE of $\mu$Q 
is obviously lower than the
corresponding NMSQE from UQ for both 
CsiNet and DualNet. The comparison clearly demonstrates one
reason why $\mu$Q can provide better CSI recovery accuracy
than UQ, as shown in our results in this section. 

\begin{table}[t]
\caption{NMSQE(dB) performance of UQ and $\mu$Q.}
\label{NMSE}
\centering
\begin{tabular}{l|c|c}
\hline
\multicolumn{1}{c|}{}     & Indoor   & Outdoor  \\ \hline
CsiNet-UQ, $M$ = 256      & -11.1237 & -12.7929 \\ \hline
CsiNet-$\mu$Q, $M$ = 256  & -19.3166 & -21.5008 \\ \hline
DualNet-UQ, $M$ = 128     & -11.33   & -7.58515 \\ \hline
DualNet-$\mu$Q, $M$ = 128 & -19.2605 & -19.5497 \\ \hline
\end{tabular}
\end{table}

\subsection{Phase Quantization}
In order to flexibly allocate the phase quantization bits under  
different bandwidth limitation and reconstruction accuracy requirement, 
we train PhaseQuan under different $\lambda$ values and illustrate
the effect of $\lambda$ on average phase quantization bits and reconstruction accuracy. Intuitively, smaller $\lambda$ leads to higher reconstruction accuracy,
though at a cost of higher feedback overhead.  The converse also holds. 



\begin{figure}[t]
\centering
\subfigure[Indoor]{%
  \includegraphics[width=0.6\columnwidth]{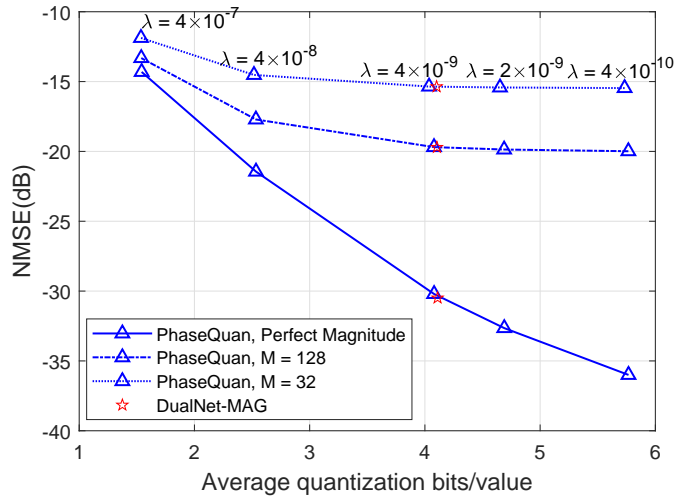}%
}\vspace*{-3mm}

\subfigure[Outdoor]{%
  \includegraphics[width=0.6\columnwidth]{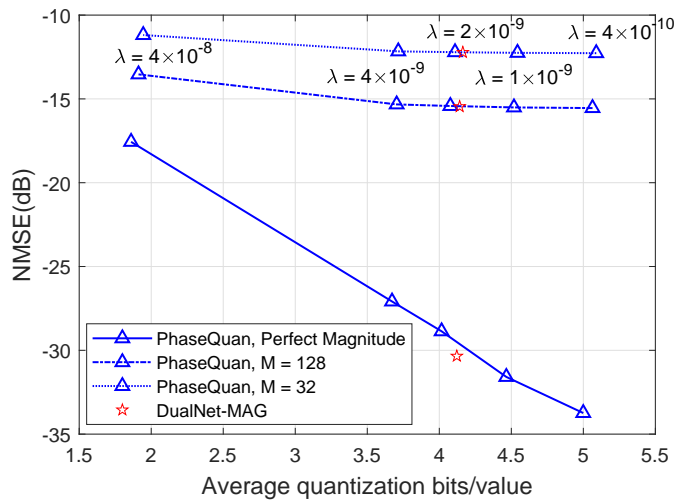}%
}\vspace*{-3mm}

\caption{Quantization bits-NMSE trade-off under different $\lambda$.}
\label{phase_Q1}
\end{figure}

The bitwidth-NMSE trade-off under different $\lambda$ is shown 
in Fig. \ref{phase_Q1}. We use the phase quantization method 
in DualNet-MAG as the baseline, and evaluate 3 cases of magnitude
knowledge for CSI reconstruction. We consider
(a) perfect CSI magnitude; (b) DualQnet after dimension compression
using 5 quantization bits
with $M = 128$, and (c) DualQnet using 5 quantization bits
with $M = 32$. 
As shown in Fig. \ref{phase_Q1}, the performances of PhaseQuan 
are comparable to the baseline with better flexibility by adjusting $\lambda$. 
NMSE decreases with increasing quantization bitwidth. 
With a large $\lambda$ value, the DNN tries to reduce the number of
quantization bits, which in turn degrades NMSE. 
To find the suitable $\lambda$ for a given NMSE, we can first select 
several candidate values of $\lambda$ to train the PhaseQuan as 
reference anchors. 
By interpolating $\lambda$ according to user’s requirements in terms of CSI
reconstruction accuracy and the available feedback bandwidth, we can
obtain the phase quantization parameters under these constraints.
 
On the other hand, with lower accuracy in magnitude, 
the influence of quantization bits becomes weaker. This means that we can 
save bits in phase quantization according to the magnitude accuracy. 
For example, in the indoor cases when $M = 32$, phase quantization 
using 2.5 bits/value and 4.1 bits/value 
can generate similar NMSE. In the outdoor cases when $M = 32$,
phase quantization using 1.9 bits/value and 4.1 bits/value 
can generate similar NMSE. 

In future works, we should jointly optimize the 
compression and encoding of magnitude and phase.

\section{Conclusions}
The previous success of DL in achieving more efficient
CSI feedback for massive MIMO systems in FDD deployment strongly
motivates investigation of bandwidth efficient encoding of the compressed CSI 
coefficients. In this paper, we propose a comprehensive DL-based CSI
feedback framework CQNet to jointly optimize the dimension compression, 
codewords quantization, and recovery of CSI matrices for massive MIMO transmission. 
We integrate CQNet with two DL-based CSI feedback mechanisms, and 
demonstrate clear feedback savings while maintaining downlink
CSI reconstruction accuracy at the massive MIMO base station. 
CQNet significantly outperforms uniform quantization and $\mu$-law quantization, 
and can reduce CSI encoding from 32 to 5 bits with little
loss of CSI reconstruction accuracy. We achieve additional feedback reduction
by introducing additional entropy encoding.
We further present a DL-based phase quantizer for CSI feedback framework DualNet-MAG
that exploits bi-directional correlation and improve
the flexibility to manage the trade-off between
CSI reconstruction accuracy and feedback bandwidth.


\ifCLASSOPTIONcaptionsoff
  \newpage
\fi

\bibliographystyle{IEEEtran}

\end{document}